\documentclass[12pt]{article}
\usepackage{epsfig}
\usepackage{amsmath}
\usepackage{hhline}
\usepackage{amssymb}
\usepackage{times}
\usepackage{cite}
\usepackage{array}
\usepackage{multirow}
\usepackage{rotating}
\usepackage{color}
\definecolor{rltred}{rgb}{0.75,0,0}
\definecolor{rltgreen}{rgb}{0,0.5,0}
\definecolor{rltblue}{rgb}{0,0,0.75}

\newlength{\dinwidth}
\newlength{\dinmargin}
\newcommand{\pom}{{I\!\!P}}
\newcommand{\reg}{{I\!\!R}}
\newcommand{\xpom}{\ensuremath{x_{I\!\!P}}}
\newcommand{\zpom}{\ensuremath{z_\pom}}
\newcommand{\gevsq}{\ensuremath{\mathrm{GeV}^2} }
\newcommand{\gev}{\ensuremath{\mathrm{GeV}} }
\newcommand{\qsq}{\ensuremath{Q^2}}

\newcommand{\pjetone}{\ensuremath{p_{jet1} }}
\newcommand{\pjettwo}{\ensuremath{p_{jet2} }}
\newcommand{\simgeq}{\ensuremath{\stackrel{>}{_\sim}}}

\begin{document}

\begin{titlepage}
\noindent
\begin{flushleft}
DESY 07-115\hfill ISSN 0418-9833\\
August 2007
\end{flushleft}

\vspace{2cm}
 
\begin{center}
\begin{Large}
 
{\bf Dijet Cross Sections and Parton Densities\\
in Diffractive DIS at HERA}

  \vspace*{1cm}
    {\Large H1 Collaboration} 
\end{Large}
\end{center}

\begin{abstract}

\noindent
%%%%%%%%%%%%%%%%%%%%%%%%%%%%%%%%%%%%%%%%%%%%%%%%%%%%%%%%%%%%%%%%%%%%%%%%%%%%%%%%%%%%%
%\input{abstract}
%%%%%%%%%%%%%%%%%%%%%%%%%%%%%%%%%%%%%%%%%%%%%%%%%%%%%%%%%%%%%%%%%%%%%%%%%%%%%%%%%%%%% 
Di{f}ferential dijet cross sections in  diffractive deep-inelastic scattering are measured with the H1 detector at HERA using an
 integrated luminosity of 51.5 pb$^{-1}$.
The selected events are of the type $ep\rightarrow eXY$, where
the system $X$ contains at least two jets %, identified
%using the inclusive $k_T$ cluster algorithm in the $\gamma^\star p $ rest frame,
 and is well separated in rapidity from the low mass  proton dissociation system $Y$.
%Dijets in diffraction are predominantly produced by the boson gluon fusion process, making the diffractive dijet cross section particularly sensitive to the diffractive gluon density.
 The dijet data are compared with QCD predictions at next-to-leading
 order based on diffractive parton distribution functions previously extracted from measurements of inclusive diffractive deep-inelastic scattering. 
The prediction describes the dijet data well  at low and intermediate $\zpom$
(the fraction of the momentum of the diffractive exchange carried by
 the parton entering the hard interaction) where the gluon density is well determined from the inclusive diffractive data, supporting QCD
factorisation. 
A new set of diffractive parton distribution functions is obtained through
 a simultaneous fit to the diffractive inclusive and dijet cross sections. 
This allows for a precise determination of both the diffractive quark and
gluon distributions in the range $0.05<\zpom<0.9$. In particular, the
precision on the gluon density at high momentum fractions is improved compared
to previous extractions. %using solely inclusive data.

\end{abstract}
\vspace{3cm}
\begin{center}
{\it Submitted to the Journal of High Energy Physics}
\end{center}
\end{titlepage}

\begin{flushleft}
%-- H1AUTS Author list by names 
%-- Status: Tue Apr  3 09:50:21 CEST 2007  Number of authors = 288 

A.~Aktas$^{11}$,               %DESY-LEFT      09/06           Aktas               
C.~Alexa$^{5}$,                %BUCH-PD        06/06           Alexa               
V.~Andreev$^{25}$,             %LPI -PD        8/88            Andreev             
T.~Anthonis$^{4}$,             %ANTW-LEFT      04/06           Anthonis            
B.~Antunovic$^{26}$,           %MPIM-ST        09/03           Antunovic           
S.~Aplin$^{11}$,               %DESY-PD        01/04           Aplin               
A.~Asmone$^{33}$,              %ROME-ST        07/2            Asmone              
A.~Astvatsatourov$^{4}$,       %BRUX-PD        07/04           Astvatsatourov      
S.~Backovic$^{30}$,            %PODG-PD        03/2            Backovic            
A.~Baghdasaryan$^{38}$
% ...

%-- H1 Institutes 
\bigskip{\it
 $ ^{1}$ I. Physikalisches Institut der RWTH, Aachen, Germany$^{ a}$ \\
% ...
$ ^{4}$ Inter-University Institute for High Energies ULB-VUB, Brussels;
          Universiteit Antwerpen, Antwerpen; Belgium$^{ c}$ \\
 $ ^{5}$ National Institute for Physics and Nuclear Engineering (NIPNE) ,
          Bucharest, Romania \\
% ...
$ ^{11}$ DESY, Hamburg, Germany \\
% ...
$ ^{25}$ Lebedev Physical Institute, Moscow, Russia$^{ e}$ \\
 $ ^{26}$ Max-Planck-Institut f\"ur Physik, M\"unchen, Germany \\
% ...
$ ^{30}$ Faculty of Science, University of Montenegro,
          Podgorica, Montenegro$^{ e}$ \\
% ...
$ ^{33}$ Dipartimento di Fisica Universit\`a di Roma Tre
          and INFN Roma~3, Roma, Italy \\
% ...
$ ^{38}$ Yerevan Physics Institute, Yerevan, Armenia \\
% ...
$ ^a$ Supported by the Bundesministerium f\"ur Bildung und Forschung, FRG,
      under contract numbers 05 H1 1GUA /1, 05 H1 1PAA /1, 05 H1 1PAB /9,
      05 H1 1PEA /6, 05 H1 1VHA /7 and 05 H1 1VHB /5 \\
% ...
$ ^c$ Supported by FNRS-FWO-Vlaanderen, IISN-IIKW and IWT
      and  by Interuniversity
Attraction Poles Programme,
      Belgian Science Policy \\
% ...
$ ^e$ Supported by the Deutsche Forschungsgemeinschaft \\
% ...
}

\end{flushleft}

\pagebreak
\pagestyle{plain}
%%%%%%%%%%%%%%%%%%%%%%%%%%%%%%%%%%%%%%%%%%%%%%%%%%%%%%%%%%%%%%%%%%%%%%%%%%%%%%%%%%%%%
%\input{introduction}
%%%%%%%%%%%%%%%%%%%%%%%%%%%%%%%%%%%%%%%%%%%%%%%%%%%%%%%%%%%%%%%%%%%%%%%%%%%%%%%%%%%%%
\section{Introduction}
Hadron-hadron collisions proceed predominantly via soft interactions to which  perturbative
Quantum Chromodynamics (QCD) cannot be applied. 
In a sizeable fraction of these soft processes the colliding  hadrons
remain intact or merely dissociate to larger mass states with the same
quantum numbers. These
 ``diffractive processes''  dominate the behaviour of the total  cross section at high energy 
and are phenomenologically described by the exchange of the pomeron trajectory,
 which carries
 the quantum numbers of the vacuum. The parton composition of this diffractive
 exchange is, however, not well known.

Processes of the type $ep\rightarrow eXp$ have been studied in detail
at HERA. These processes can be pictured as $\gamma^\star p$
scattering, where the virtual photon interacts with a diffractive
exchange and dissociates to produce a system $X$.
In QCD a hard scattering collinear factorisation theorem~\cite{Collins}
predicts that the cross section for diffractive deep-inelastic $ep$
scattering (DIS) factorises into a set of universal diffractive parton
distribution functions (DPDFs) of the proton and process-dependent hard
scattering coefficients. DPDFs have been
determined through QCD fits to the measured
cross sections of inclusive diffractive scattering at HERA~\cite{h1f2d94,Martin1,Martin2,h1f2d97,Chekanov:2004hy,Hautmann:1999ui,Hautmann:2000pw,Breitweg:1998fn,Golec-Biernat:1995bn,Gehrmann:1995by,Alvero:1998ta,Royon:2000wg,Watt:2005gt,Martin:2006td}.

\begin{figure}[h!]
\begin{center}
\begin{minipage}{.5\textwidth}
\includegraphics[width=\textwidth]{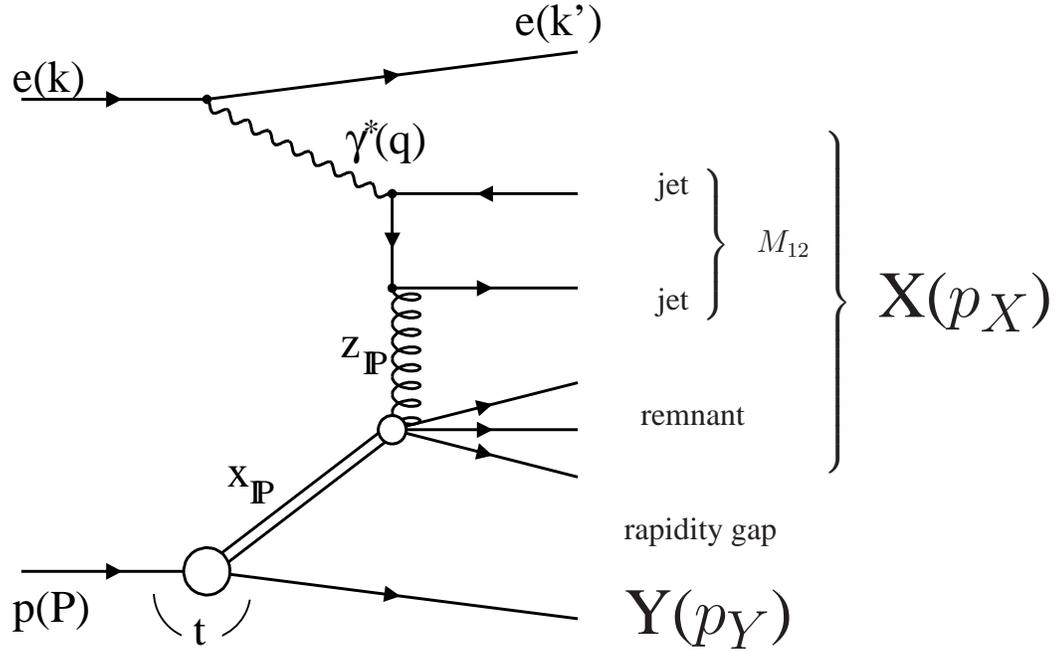}
\end{minipage}
\begin{minipage}{.3\textwidth}{\vspace{3.0cm}$\begin{array}{ll}\left. \begin{array}{ll} &  \\ \left. \begin{array}{l} \text{jet} \\ \\ \\ \text{jet} \\  \end{array} \right\} &  M_{12}\\ & \\ & \\ \multicolumn{2}{l}{\text{remnant}} \\ & \\ \end{array} \right\} & \parbox[c]{1cm}{\Huge{X($p_X$)}}\\
 & \\ \text{rapidity gap}& \\  & \\ \parbox[c]{1cm}{\Huge{Y($p_Y$)}} & \\ & \\ &\end{array} $}
\end{minipage}
\caption{Leading order diagram for diffractive dijet production in DIS.}
          \label{feynman}
        \end{center}
      \end{figure}

If QCD factorisation is fulfilled, next-to-leading order (NLO) QCD
calculations based on DPDFs such as those extracted in~\cite{h1f2d97}
should be able to predict the production rates of more exclusive
diffractive processes such as dijet and open charm production. 
Previous  measurements of such exclusive cross sections  in 
 DIS~\cite{disjets1,disjets2,disjets3,charm1,charm2,Chekanov:2003gt,Chekanov:2002qm,Chekanov:2001cm} support QCD factorisation since they  can be reasonably well described using the DPDFs determined from  inclusive diffractive scattering. Diffractive dijet and charm production proceed mainly via  boson gluon fusion (BGF, depicted in figure~\ref{feynman}) and are therefore mainly sensitive to the diffractive gluon density. It was recently shown~\cite{h1f2d97} that inclusive diffractive scattering data do not constrain the diffractive gluon density well at high momentum fractions. Thus stringent tests of factorisation can only be performed at low momentum fractions. However, the gluon density at high momentum fractions is particularly relevant for the estimation of cross sections for several important processes at the LHC~\cite{khoze}.
%However, the phase space covered by measurements of diffractive dijet and charm production at HERA falls largely in a kinematic region where the diffractive gluon density is poorly determined  from inclusive diffractive scattering~\cite{h1f2d97}.
%Thus, on the one hand, QCD factorization can only be stringently tested in a limited kinematic domain, as both diffractive dijet and charm production are expected mainly to proceed via the process of boson gluon fusion (BGF) as depicted in figure~\ref{feynman} and NLO QCD predictions of their cross sections are mainly determined by the diffractive gluon density.
%%%
%%%%%%
%=======================
Measurements of diffractive dijet production can directly
constrain the diffractive gluon density at high momentum fractions, extending the kinematic range of reliably determined diffractive parton densities.

  In this paper, a new  measurement of diffractive dijet cross sections
in deep-inelastic scattering  is presented, based on data collected with
the
H1 detector at HERA in the years 1999 and 2000. 
These are the first HERA diffractive DIS dijet data with
$E_p=920~\gev$.
Jets are defined using the inclusive $k_T$ 
algorithm~\cite{kt}. % with
%asymmetric cuts on the jet transverse energies to facilitate
%comparisons with next-to-leading order predictions. 
The resulting dijet cross sections are compared to NLO QCD predictions 
 based on DPDFs previously extracted~\cite{h1f2d97} from inclusive diffractive
 $ep$ scattering at H1. For the first time, a combined
 NLO QCD fit is performed to the differential dijet cross sections and the
inclusive diffractive cross section data in order to determine a new
 set of DPDFs.% with better precision in particular the high $\zpom$ region for the gluon density.

%%%%%%%%%%%%%%%%%%%%%%%%%%%%%%%%%%%%%%%%%%%%%%%%%%%%%%%%%%%%%%%%%%%%%%%%%%%%%%%%%%%%%
%\input{kinematics}
%%%%%%%%%%%%%%%%%%%%%%%%%%%%%%%%%%%%%%%%%%%%%%%%%%%%%%%%%%%%%%%%%%%%%%%%%%%%%%%%%%%%%
\section{Kinematics} 
The dominant process leading to the production of dijets in diffractive DIS is depicted in \mbox{figure~\ref{feynman}}. The incoming proton of four-momentum $P$ interacts with the positron of four-momen\-tum $k$ via the exchange of  a virtual photon with four-momentum $q$. The DIS kinematic variables are defined as 
 
\begin{equation*}
 \qsq \equiv  - q^2, \qquad  x \equiv \frac{-q^2}{2P \cdot q}, \qquad y \equiv \frac{P \cdot q}{P \cdot k},
\end{equation*} 
where $\qsq$ is the photon virtuality, $x$ is the longitudinal momentum fraction of the proton carried  by the struck quark and $y$ is the  inelasticity of the process. These quantities are connected by the relation

\begin{equation*}
Q^2=x y  s,
\end{equation*}
where $s$ denotes the fixed $ep$ centre-of-mass energy squared.

The hadronic final state  of the events is divided into two systems $X$ and $Y$, separated
 by the largest  gap in the rapidity distribution  of the hadrons relative to the collision axis in the $\gamma^\star p $
 centre of mass system.
The diffractive scattering is described in terms of the variables

\begin{equation*}
%M^2_X\equiv p_X^2; \quad M^2_Y\equiv p_Y^2; \quad 
t\equiv (P-p_Y)^2,\quad
\xpom \equiv \frac{q\cdot (P-p_Y)}{q\cdot P},\quad \beta\equiv x/\xpom,
\end{equation*}
with $p_Y$ representing the four-momentum of the system $Y$.
Here %$M_X$ and $M_Y$ denote the invariant masses of the systems $X$ and $Y$, 
$t$ is the squared four-momentum transfer at the proton vertex and $\xpom$ is the fraction of the proton's longitudinal momentum
transferred to the system $X$. 
The fractional  longitudinal  momentum of  the diffractive exchange carried by the parton 
which enters the hard interaction with
 four-momentum $v$ is given by

\begin{equation*}
 \zpom=\frac{q\cdot v}{q\cdot
(P-p_Y)}.
\end{equation*}
%where $v$ is the four-momentum of the parton from
% the diffractive exchange which enters the hard interaction. 

%\input{expproc}
%%%%%%%%%%%%%%%%%%%%%%%%%%%%%%%%%%%%%%%%%%%%%%%%%%%%%%%%%%%%%%%%%%%%%%%%%%%%%%%%%%%%%
\section{Experimental Procedure}
\subsection{H1 Detector}
\label{detector}
A detailed description of the H1 detector can be found in \cite{H1det1,H1det2,spacal}.
Here, a brief account of the components most relevant to the present 
analysis is given.
The origin of the H1 coordinate system is the nominal
$ep$ interaction point. The direction of the proton beam defines
the positive $z$--axis (forward direction). Transverse momenta are measured in the $x$--$y$ plane. Polar~($\theta$) and~azimuthal~($\phi$) angles are measured with respect to this reference system.
The pseudorapidity is defined as $\eta= -\ln{\tan (\theta/2)}$.

%The hadronic final state $X$ is measured with a tracking and a
%calorimeter system.
 The $ep$ interaction region is surrounded by a two-layered silicon strip detector~\cite{cst}
and two large concentric drift
chambers, operated inside a 1.16~T solenoidal magnetic field. Charged particle momenta are measured
 in the pseudorapidity range
$-1.5< \eta <1.5$ with a resolution of $\sigma(p_T)/p_T=0.005\,
 p_T/$GeV$\, \oplus \,$0.015. 
The central tracking detectors also provide triggering
information based on track segments measured in the $r$-$\phi$ plane of the
central jet chambers and on the $z$ position of the event vertex obtained from the double
layers of two multi-wire proportional chambers.
A finely segmented electromagnetic
 and hadronic liquid argon (LAr) calorimeter~\cite{Andrieu:1993kh} covers the
range $-1.5 < \eta < 3.4$. The energy resolution is
$\sigma/E=0.11/\sqrt{E/{\rm GeV}}$
 for electromagnetic showers and $\sigma/E=0.50/\sqrt{E/{\rm GeV}}$ for hadrons, as
measured in test beams~\cite{lartestbeam}. A lead/scintillating fibre calorimeter (SPACAL)~\cite{spacal} covers the backward region $-4 < \eta < -1.4$. Its main purpose is the detection of scattered positrons.

The luminosity is measured via the Bethe-Heitler Bremsstrahlung
process $ep \rightarrow ep \gamma$, the final state photon being detected in a
crystal calorimeter at $z=-103$ m. 

The Forward Muon Detector (FMD) and the
Proton Remnant Tagger (PRT) are sensitive to the energy flow in the
forward region. % , which allow an efficient rapidity gap selection.
They are used to efficiently reject events which do not
exhibit a  rapidity gap between the $X$ system and the proton
dissociation system $Y$.
The FMD is located at
$z=6.5~{\rm m}$ and covers a pseudorapidity range of $1.9< \eta <3.7$\@. It
may also detect particles produced at larger $\eta$ due to secondary scattering
within the beam pipe. 
 The PRT
consists of a set of scintillators surrounding the beam
pipe at $z=26$ m and covers the
region $6 < \eta < 7.5$.

\subsection{Event Selection}
The data used in this analysis correspond to an integrated luminosity of 51.5~pb$^{-1}$  taken in the 1999 and 2000 running
periods, in which HERA collided protons of  920~GeV energy with
positrons of  27.5~GeV. The data are collected using a trigger which requires the 
scattered positron to be detected in the SPACAL calorimeter
and at least one track of transverse momentum above $0.8~\gev$ to be recorded
in the central jet chamber. In the off-line analysis, the scattered
positron is selected as an electromagnetic SPACAL cluster with an
energy  $E_{e}  > 8$~GeV and polar angle \mbox{$156^\circ < \theta_e < 176^\circ$}. 
%The  kinematic range is chosen to be 
These requirements are well matched to the chosen kinematic range of
 $ 4 < Q^2 < 80~\gevsq$ and $0.1 < y < 0.7$. Background  from
 photoproduction, where the positron scatters unobserved
at
small angles and a particle from the hadronic final state is misidentified
 as the scattered positron, is suppressed by the requirement that the
 difference between the total energy and  longitudinal momentum
 reconstructed in the detector, $E-p_z$, must be larger than
 $35~\gev$. Background not related to $ep$ collisions is reduced by
 restricting the $z$ position of the event vertex to lie within $35$~cm of the average $ep$ interaction point. 

Diffractive events are selected by the absence of hadronic activity
 above noise threshold in the most forward part of the LAr calorimeter
 ($\eta > 3.2$) and in the FMD and PRT. This selection ensures
 that the rapidity gap between the systems $X$ and $Y$ spans more than
 four units between $\eta=3.2$ and $7.5$.  
In addition the restriction $\xpom < 0.03$ is imposed to limit the
  contribution from secondary reggeon exchanges and to ensure
 good acceptance.

The hadronic system $X$ is measured in the LAr and SPACAL calorimeters and the
central tracking system. Calorimeter cluster energies and track momenta are
combined into hadronic objects using an algorithm which avoids double
counting~\cite{combobj}.  Jets are formed from the hadronic objects, using the
inclusive $k_T$ cluster algorithm~\cite{kt} with a distance parameter of unity
in the photon-proton rest frame.
At least two jets are required with transverse momenta in the
 $\gamma^\star p$ centre of mass frame of $p_{T,jet1}^\star  > 5.5$ GeV
and $p_{T,jet2}^\star  > 4$ GeV for the   leading and sub-leading jet,
respectively. Asymmetric cuts on the jet transverse momenta are
 chosen to facilitate comparisons with NLO QCD predictions. 
The  axes of the jets  are required to lie within the region $-1.0 < \eta_{jet} < 2.0$ in the laboratory frame, well within
the acceptance of the LAr calorimeter. %(the cross section at  hadron level is defined for a closely corresponding rapidity region in the $\gamma^\star p$ system). 
 After all cuts 2723 diffractive dijet events are selected. 
% This sample is a factor of seven larger than the one used in the earlier study~\cite{disjets3}.

%%%%%%%%%%%%%%%%%%%%%%%%%%%%%%%%%%%%%%%%%%%%%%%%%%%%%%%%%%%%%%%%%%%%%%%%%%%%%%%%%%%%%
%\input{reconstruction}
%%%%%%%%%%%%%%%%%%%%%%%%%%%%%%%%%%%%%%%%%%%%%%%%%%%%%%%%%%%%%%%%%%%%%%%%%%%%%%%%%%%%%
\subsection{Kinematic Reconstruction}
The energy $E_{e}$ and polar angle $\theta_e$ of the scattered positron are
measured using the SPACAL and the reconstructed vertex position. The
inelasticity $y$ and photon virtuality $\qsq$ are determined according to
\begin{xalignat*}{2}
y    &=1-\frac{E_{e}}{E^0_e}\sin^2\frac{\theta_e}{2},\\
\qsq &=4E_{e}E^0_e\cos^2\frac{\theta_e}{2},
\intertext{%\end{xalignat*}
where $E^0_e$ is the positron beam energy. 
The energy and momentum of the hadronic system $X$
are reconstructed from the observed hadronic objects
 and the invariant mass $M_X$ is computed from this information. 
 The invariant mass of the dijet system is given by
}%\begin{equation*}
M_{12} &\equiv \sqrt{\strut \left( \pjetone + \pjettwo \right)^2},
\intertext{%\end{equation*}
with $\pjetone$ and $\pjettwo$ being the four-momenta of the leading and
sub-leading jet, respectively.
The observables $\xpom$ and $\zpom$ are reconstructed according to
}%\begin{xalignat*}{2}
     \xpom &= \frac{M_X^2+\qsq}{ys},\\
     \zpom &= \frac{M_{12}^2+\qsq}{M_X^2+\qsq}.
\end{xalignat*}

\subsection{Monte Carlo Simulations and Fixed Order QCD Predictions}
Monte Carlo simulations are used
in the analysis to correct the data for detector effects. 
For events generated with Monte Carlo programs, the H1 detector
response is simulated in detail using GEANT~\cite{geant_manual} and the events are subjected to the same analysis as the data.
Events are generated using the RAPGAP program~\cite{RAPGAP}
which simulates the process $ep\rightarrow eXp$, assuming proton
vertex factorisation (see section~\ref{fit_ansatz}).
Leading order matrix elements for the hard QCD sub-process are
convoluted with DPDFs, taken at the factorisation scale $\mu_f=\sqrt{\hat{p}_T^2+\qsq}$,
where $\hat{p}_T$ is the transverse momentum of the emerging hard
partons relative to the collision axis in the $\gamma^\star p$ centre of mass frame.
A preliminary version of the `H1 2006 DPDF' fit~\cite{fit2002} is
used to simulate pomeron and sub-leading reggeon exchanges.
Higher order effects are simulated using parton showers~\cite{PS} in
the leading logarithm approximation. The Lund string model
\cite{LUND1,LUND2} is used for hadronisation.  QED radiative
corrections are  applied using the HERACLES
program~\cite{heracles}. Processes with a resolved virtual photon  are
also   included, with the structure of the photon given by the SAS-2D
parameterisation \cite{SAS-2D}.

The background due to non-diffractive deep-inelastic scattering is
estimated and accounted for using the RAPGAP Monte Carlo program 
in its 
inclusive mode. The parameters are chosen to be similar to the ones
used for the generation of the 
diffractive sample discussed above. The inclusive simulation uses the
CTEQ5L parton densities of the proton~\cite{Lai:1999wy}.

In diffractive DIS measurements using the present technique the system
$Y$ does not necessarily consist only of a single proton, but may also
be a low mass dissociative system. 
The DIFFVM program~\cite{DIFFVM} includes a sophisticated treatment of
the dissociating proton. It is used to study the response of the forward detectors to low  mass proton dissociation systems
($m_p<M_Y<5~\gev$). The non-resonant part
of the $M_Y$ distribution is modelled with 
$d\sigma / d M_Y^2\propto(1/M_Y^2)^{1.08}$, while the $t$ dependence
follows  an exponential decrease: $d\sigma / d t \propto e^{bt}$
with $b=1.6~\gev^{-2}$. This parameterisation is motivated by
measurements of diffractive vector meson production at
H1~\cite{diffvm_1}. Proton dissociation processes with $M_Y>5~\gev$
are included in the treatment of non-diffractive background with
RAPGAP as discussed above.

NLO QCD predictions for the dijet cross sections are calculated at the parton level using
 the NLOJET++  program~\cite{nlojet} in slices of $\xpom$, assuming proton vertex factorisation. 
The resulting cross sections are converted to the
 stable hadron level by  factors extracted from the 
 RAPGAP Monte Carlo model in the diffractive mode. 
 The renormalisation and factorisation
 scales are set to $\mu_r=\mu_f=\sqrt{p^{\star 2}_{T,jet1}+\qsq}$, where
 $p^\star_{T,jet1}$ is the transverse momentum of the leading jet
 in the $\gamma^\star p$ centre of mass frame. The NLOJET++ calculation uses parton densities obtained from a NLO QCD
 analysis of inclusive diffractive scattering at
 H1~\cite{h1f2d97}. That publication provides two sets of parton densities,
 H1 2006 DPDF fit A and fit B, which differ  in the parameterisation of
 the gluon density. A steeper fall-off in the
 gluon density at high $\zpom$\ is obtained for fit B than for fit A, while the quark densities agree within the uncertainties. Both DPDF sets provide a good description of the inclusive diffractive DIS data.

%The uncertainties on the parton densities are propagated to the dijet
%prediction by generating additional predictions in which 
The experimental and theoretical uncertainties on the DPDFs
in~\cite{h1f2d97} are propagated to the dijet
prediction  via an eigenvector decomposition of the error sources
according to the method presented in~\cite{Giele:2002hx}.
 The deviations from the nominal prediction are added in quadrature to obtain the uncertainty on the dijet prediction due to DPDF uncertainties. Alternative hadronisation corrections are
 extracted from the POMWIG Monte Carlo model~\cite{Cox:2000jt}, which
 uses cluster fragmentation~\cite{Marchesini:1987cf,Webber:1983if} to
 describe hadronisation. The difference between the nominal and
 alternative hadronisation corrections is taken to be the
 hadronisation uncertainty on the NLO QCD prediction. To account for
 the uncertainty due to the missing higher orders in the calculation,
 the renormalisation and factorisation scales are varied by common factors of 2 and 0.5 with respect to the nominal prediction.

%%%%%%%%%%%%%%%%%%%%%%%%%%%%%%%%%%%%%%%%%%%%%%%%%%%%%%%%%%%%%%%%%%%%%%%%%%%%%%%%%%%%%
%\input{xsmeasurement}
%%%%%%%%%%%%%%%%%%%%%%%%%%%%%%%%%%%%%%%%%%%%%%%%%%%%%%%%%%%%%%%%%%%%%%%%%%%%%%%%%%%%%
\subsection{Cross Section Measurement}
%\vspace{0.5cm}
\setlength{\tabcolsep}{0.5cm}
\begin{table}                                                                  
\begin{center}
%{\textbf{Cross Section Definition}}\\[1em] 
\begin{tabular}{|m{3cm}r@{\hspace{3mm}}c@{}c@{}c@{\hspace{3mm}}l|}
\hline
\multirow{4}{*}{\bf{DIS Selection}}&&&&&  \\
&$4 $&$<$&$ \qsq $&$<$& $80~\gevsq$\\
&$0.1      $&$<$&$ y   $&$<$& $0.7$\\
&&&&&  \\
\hline
\multirow{5}{*}{\bf{Diffractive Selection}}&&&&&  \\
&           &   &$\xpom$&$<$& 0.03\\
&           &   &$ M_Y $&$<$&$1.6~\gev $\\
&           &   &$ |t| $&$<$&$1~\gevsq$\\
&&&&&  \\
\hline
\multirow{5}{*}{\bf{Jet Selection}}&&&&&  \\
&           &   &$ p_{T,jet1}^\star $&$>$&$5.5~\gev$\\
&           &   &$ p_{T,jet2}^\star $&$>$&$4~\gev   $\\
&-3         &$<$&$ \eta^\star_{jet}     $&$<$& 0\\
&&&&&  \\
\hline
\end{tabular}

\caption[Definition of Cross Sections]{The kinematic domain in which the
cross sections are measured at the level of jets of stable hadrons. The jets are reconstructed using the
inclusive $k_T$ algorithm as described in the text. Variables marked
with a $\star$ are evaluated relative to the collision axis in the $\gamma^\star p$ centre of mass frame.}
\label{tab:xsdef}
\end{center}
\end{table}
%\vspace{0.5cm}

The measured differential dijet cross sections are defined at the level
of stable hadrons in the kinematic region specified in
table~\ref{tab:xsdef}. A correction of typically 20\% is applied to
account for detector acceptances, inefficiencies
and migrations between measurement bins using the 
RAPGAP 3.1  Monte Carlo program. This  simulation gives a reasonable
description of the shapes of all data  distributions. According to the
simulation, the detector
level observables are found to be well correlated with the observables
at 
 hadron level. The cross sections are corrected to
the QED Born level using the HERACLES interface to the
RAPGAP Monte Carlo program. The small background contribution from non-diffractive deep-inelastic scattering is statistically subtracted using the Monte Carlo sample introduced above.

%As the forward detector selection cannot guarantee an elastically
%scattered proton but only a low mass dissociative system, 
The cross
section definition for this study is chosen  to include all events with $M_Y<1.6~\gev$ and $|t|<1~\gevsq$ as in~\cite{h1f2d94,h1f2d97,disjets2,disjets1,disjets3}. As $M_Y$ and $|t|$ are not measured directly, the effects of migration across these boundaries  must be estimated.
Migrations from large $M_Y>5$~\gev  and $\xpom>0.2$  are corrected for using
RAPGAP in inclusive mode.
Smearing across the $M_Y = 1.6~\gev$ boundary of events with $M_Y\leq 5$~\gev is evaluated
with the DIFFVM~\cite{DIFFVM} simulation of proton dissociation, following~\cite{h1f2d94}.

\subsection{Systematic Uncertainties}
\label{syst_unc}
The systematic uncertainties are evaluated separately for each measurement bin, except for uncertainties on global correction factors.
The following sources of uncertainty are determined to be largely correlated between bins:
\begin{description}
\item[LAr calorimeter energy scale:] The energy scale of the LAr
  calorimeter response to hadrons is varied by $\pm4\%$ in the simulation, which causes a
  variation of the total cross section by $^{+5}_{-3}\%$ and slightly
  larger uncertainties in individual measurement bins. 
\item[Track Momenta:] The contribution of the track momenta to the
  $X$ system is varied by $\pm 3\%$, resulting in a total cross section
  uncertainty of around 3\%.
\item[Luminosity:] The measurement of the integrated luminosity has an uncertainty of 1.5\%. This translates directly into a 1.5\% uncertainty on the cross section.
\item[FMD noise:] The cross section is corrected for the fraction of
  events rejected due to noise in the
  forward muon detector. A global correction factor is
  determined from a sample of randomly triggered events and is found to be $(1.2\pm0.4)\%$. The uncertainty on this correction factor leads to  an overall normalisation uncertainty of $0.4\%$.
% This leads to an overall normalisation uncertainty of $0.4\%$.
\item[\boldmath\xpom-migration:] %The correction for migrations of
The estimated number of non-diffractive background events which
migrate into the sample from the unmeasured region $\xpom >0.03$ or
$M_Y >5~\gev$ is varied by $\pm 50\%$, leading to a total cross section
uncertainty of 1\%.
\item[\boldmath$M_Y$ and \boldmath$|t|$ migrations:] The systematic
  uncertainties connected to migrations over the $M_Y$ and $|t|$
  limits are assessed following the method of~\cite{h1f2d97}, giving a total uncertainty of 5\%.
\item[Rapidity gap selection inefficiency:] A fraction of the events
  in the kinematic range specified in table~\ref{tab:xsdef} give rise
  to hadronic activity at pseudorapidities larger than allowed by the $\eta_{max}$ cut in the LAr calorimeter or in the forward detectors and is thus lost. The correction for this effect relies heavily on the RAPGAP
simulation to describe the forward energy flow of diffractive
  events. The forward energy flow in diffractive DIS is investigated
  with a sample of elastically scattered protons  detected in the
  forward proton spectrometer of the H1 detector~\cite{FPS}. The study finds the
  RAPGAP model to describe these migrations to within $30\%$~\cite{schenk}. The effect
  of this uncertainty on this measurement is estimated by reweighting
  all events in the signal simulation which do not pass the forward
  detector cuts by $\pm 30\%$. This translates into an uncertainty of
  $^{+10}_{-5}\%$ on the total cross section.
\end{description}
The remaining systematic uncertainties, described below, show
significantly less correlation and are thus treated as uncorrelated
between measurement bins.
\begin{description}
\item[Positron energy:] The energy of the scattered positron is known
  to within 2\% at $E_{e}=8~\gev$, falling linearly to 0.3\% at
  $E_{e}=27.5~\gev$. This translates  into a 2\% uncertainty on the total cross section.
\item[Positron angle:] The uncertainty in the polar angle $\theta_e$
  of the scattered positron is 1~mrad. This contributes an uncertainty
  of $1\%$ to the total cross section. 
\item[Trigger efficiency:] The average difference between the trigger
  efficiency as extracted from the Monte Carlo simulation and from the
  data using monitor trigger samples is taken as the uncertainty on
  the trigger efficiency, which is  around 1\%.
\item[Unfolding uncertainties:] To evaluate the model dependence of
  the correction from the detector to the hadron level, key kinematic dependences of the Monte Carlo simulation are reweighted within the limits imposed by the present data. The following distributions are varied: 
\xpom\ by $\xpom^{\pm 0.2}$, 
$\hat{p}_T$ by $\hat{p}_T^{\pm 0.4}$, 
$|t|$ by $e^{\pm2t~\gev ^{-2}}$ and  
$y$ by $y^{\pm0.3}$. 
The largest uncertainty is introduced by the $\hat{p}_T$
  reweighting (typically 4\%) followed by 
\xpom\ (3\%), while the reweights in the two other variables have rather small effects. 
\end{description}

The largest contributions to the  systematic errors on the cross sections arise from the 
uncertainty in the LAr calorimeter energy scale, from unfolding
uncertainties and from the rapidity gap selection inefficiency.
 The overall uncertainty on the total cross section is $^{+15}_{-10}\%$. The uncertainties on
 individual measurement bins are slightly larger.% and are listed in the

\section{Dijet Results}
The integrated cross section in the kinematic range specified in table~\ref{tab:xsdef} is determined to be
\begin{equation*}
\sigma^{2jets}(ep\rightarrow eXY)=52\pm 1~(\mathrm{stat.})~^{+7}_{-5}~(\mathrm{syst.})~\mathrm{pb}.
\end{equation*}
When this measurement is translated to the kinematic range of the
previous H1 result~\cite{disjets3} (i.e. after correcting for the
different proton beam energies, $y$-range and
$p^\star_{T,jet1}$-ranges), the two results are compatible within
the uncertainties. The total cross section can be compared to the
NLO QCD predictions based on the two sets of DPDFs determined from inclusive diffraction~\cite{h1f2d97}:
\begin{xalignat*}{2}
\sigma^{2jets}(\mathrm{H1\ 2006\ DPDF\ fit\ A})&=75~^{+27}_{-17}~(\mathrm{scale\ unc.})~\pm 7(\mathrm{DPDF})~\mathrm{pb},\\
\sigma^{2jets}(\mathrm{H1\ 2006\ DPDF\ fit\ B})&=57~^{+21}_{-13}~(\mathrm{scale\ unc.})~\pm 8(\mathrm{DPDF})~\mathrm{pb}.
\end{xalignat*}
The scale uncertainty is derived by simultaneously  varying $\mu_f$
 and $\mu_r$  by common  factors of 2 and 0.5.
% The H1 2006 DPDF fit A overestimates the total cross section by $\sim 40\%$, while the prediction of the H1 2006 DPDF fit B for the total cross section is compatible with the measurement within the experimental uncertainties.
Whilst both predictions are compatible
with the measurement, the central result of fit A overestimates the
cross section by $\sim40\%$.

\begin{figure}[h!]
\hspace{1cm}
\includegraphics[scale=.6]{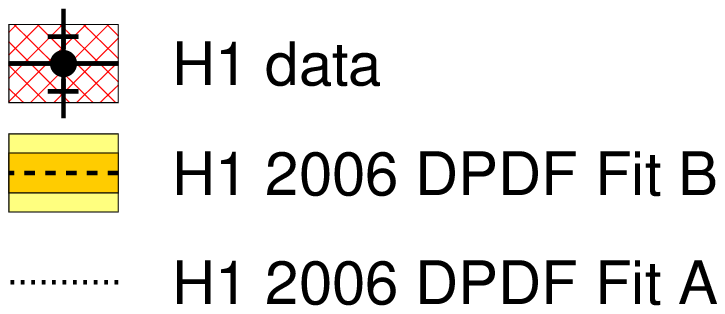}
\begin{center}
\includegraphics[width=.45\textwidth]{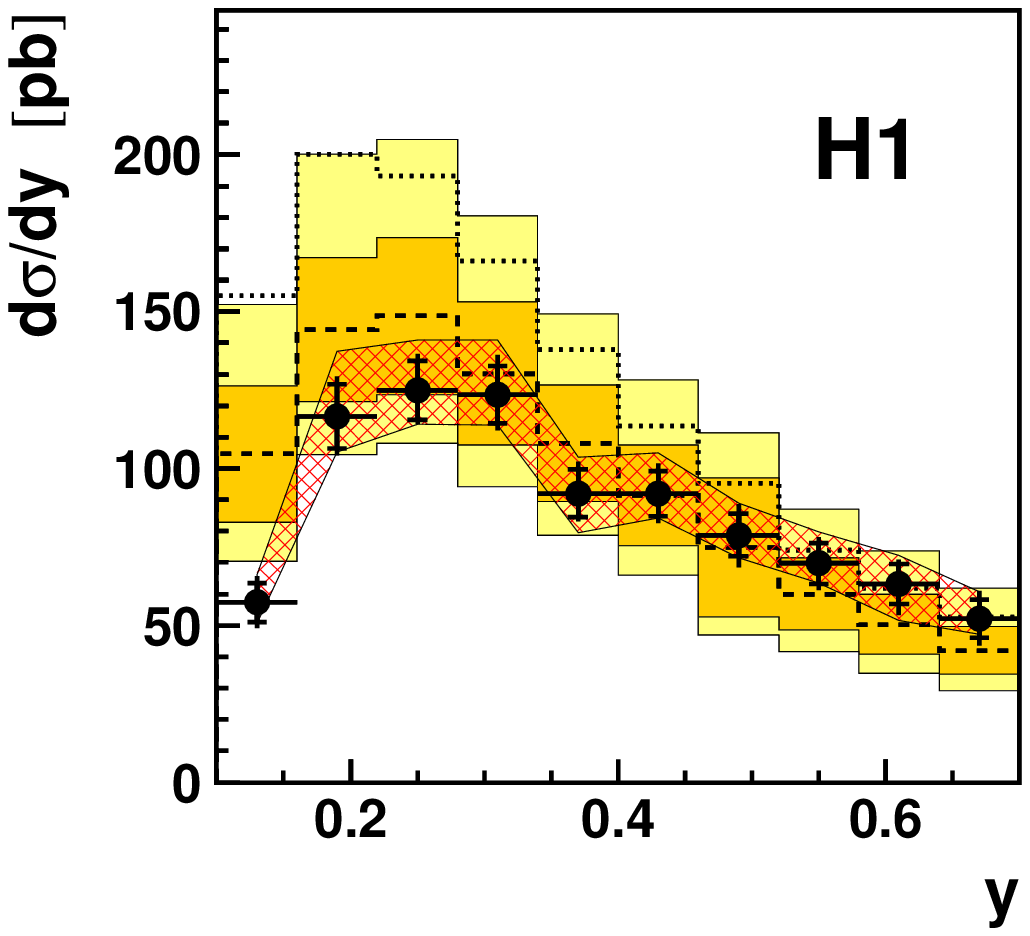}
\hspace{.05\textwidth}
\includegraphics[width=.45\textwidth]{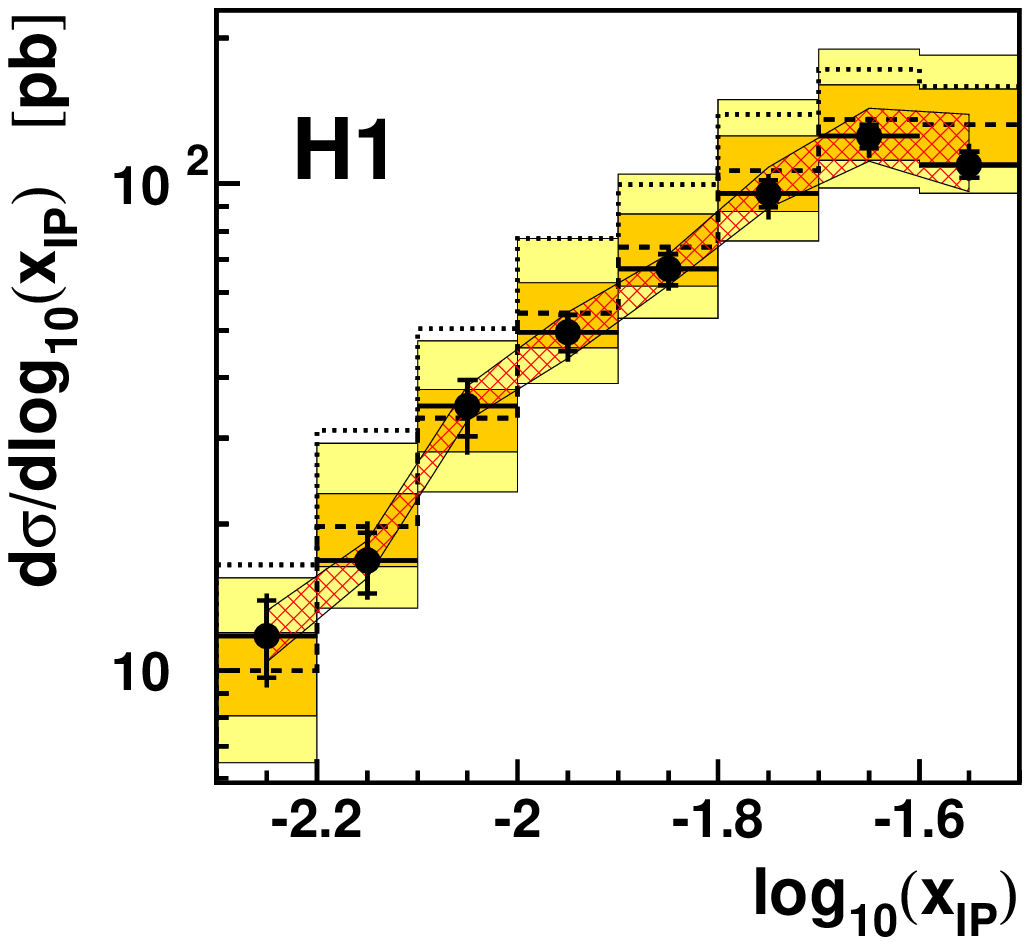}
\vspace{.05\textwidth}\\
\includegraphics[width=.45\textwidth]{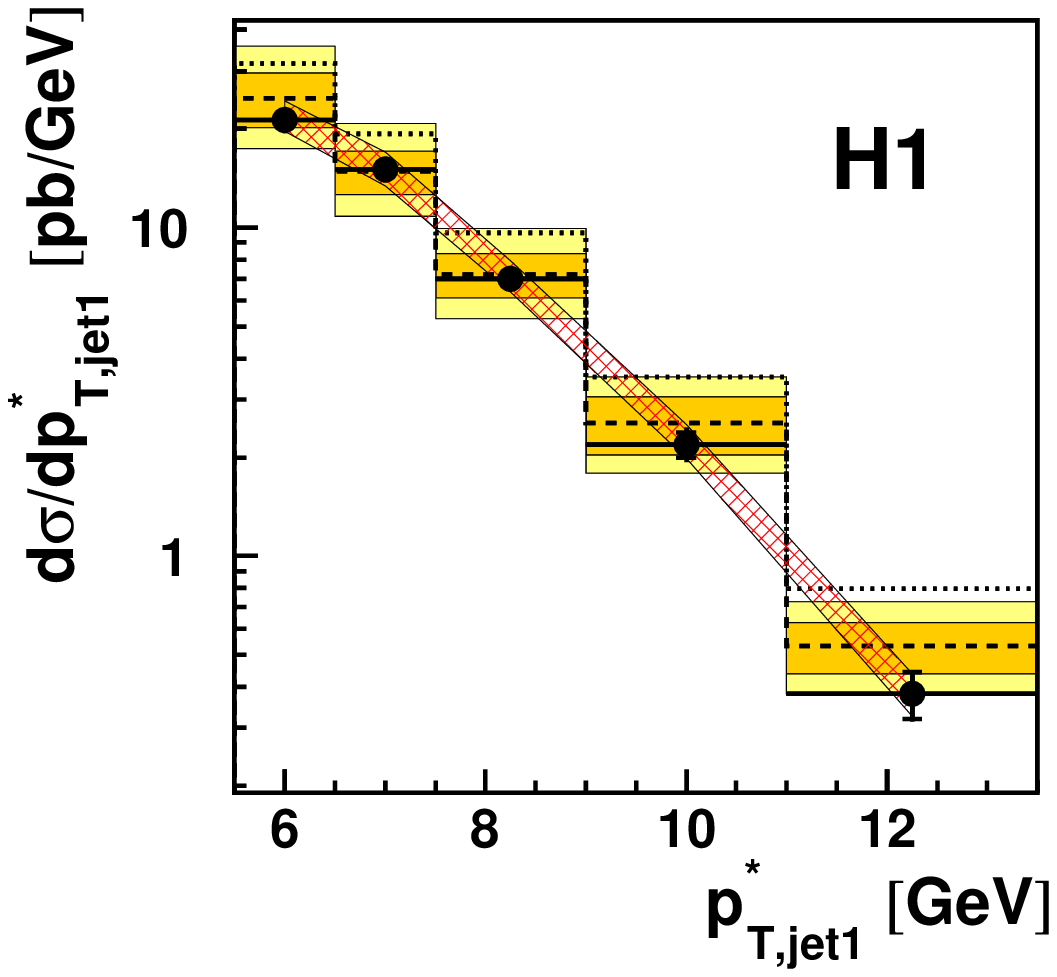}
\hspace{.05\textwidth}
\includegraphics[width=.45\textwidth]{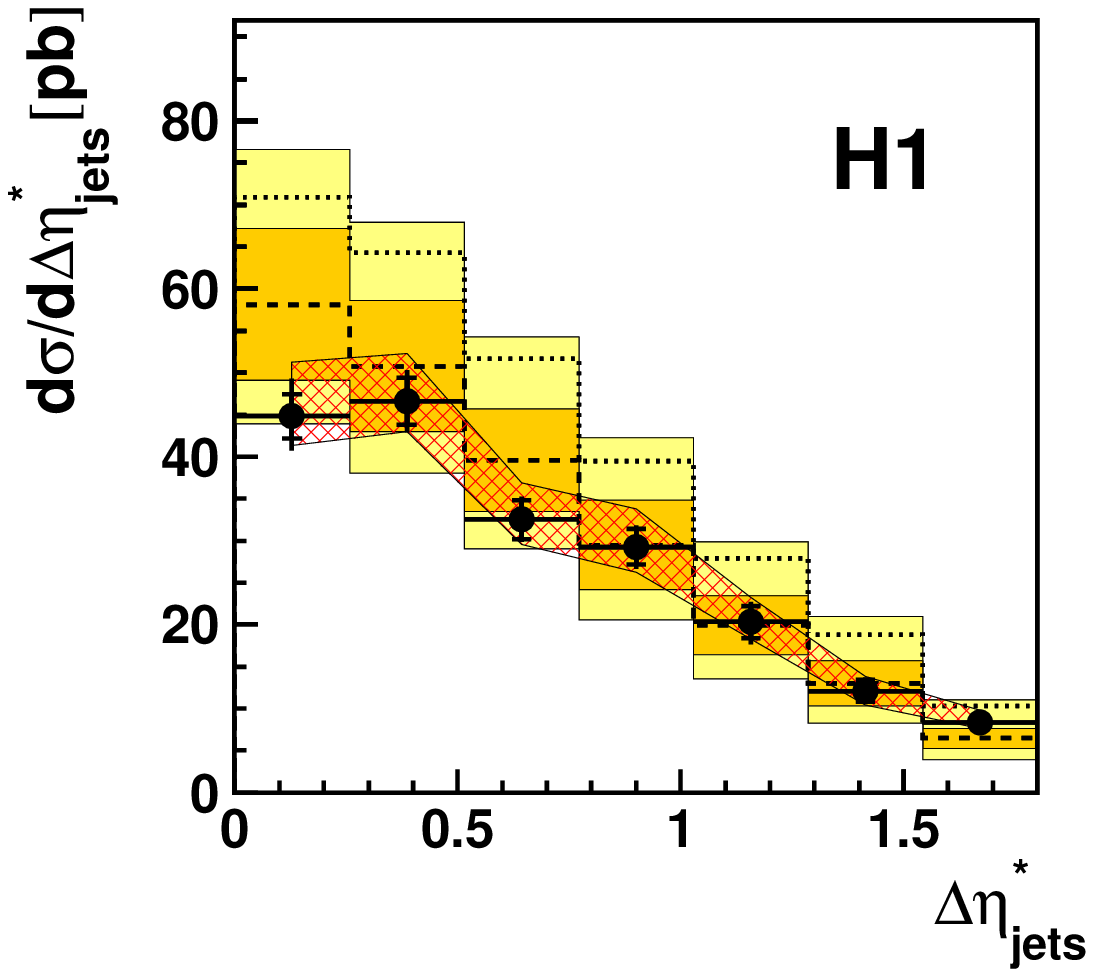}
\caption{Cross sections for diffractive dijets, differential in  $y$,
$\log\xpom$, $p^\star_{T,jet1}$ and $\Delta\eta^{\star}_{jets}$ compared to NLO
  predictions based on the parton-densities from the H1 2006 DPDF
  fits~\cite{h1f2d97}. The data are shown as black points with the
  inner and outer error bars denoting the statistical and quadratically added uncorrelated
  systematic uncertainties, respectively. The hatched band
  indicates the correlated systematic uncertainty. The dashed
  line shows the NLO QCD prediction based on the H1 2006 DPDF fit B, which is 
  surrounded by a dark shaded band indicating the parton density and hadronisation
  uncertainties. In the light shaded band the scale uncertainty is added
  quadratically to the parton density and hadronisation
  uncertainties. The
  dotted line represents the NLO QCD prediction based
  on the H1 2006 DPDF fit A.}
\label{fig:kin_fp}
\end{center}
\end{figure}

\begin{figure}[h!]
\begin{center}
\begin{minipage}{0.45\textwidth}
\begin{flushleft}
\hspace{1cm}
\includegraphics[scale=.6]{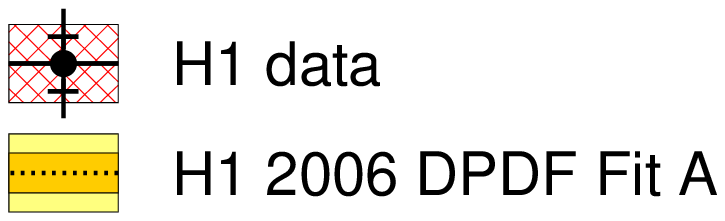}
\end{flushleft}
\includegraphics[width=\textwidth]{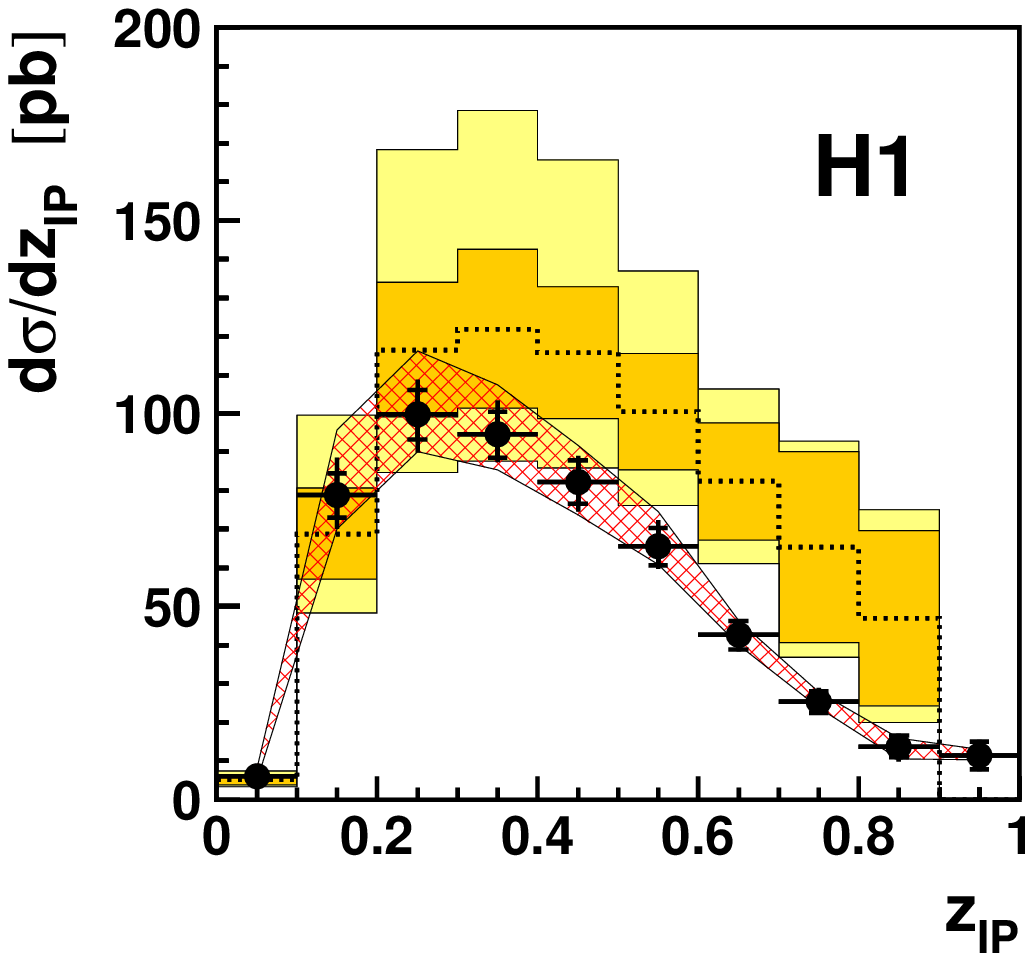}
\end{minipage}
\hspace{0.05\textwidth}
\begin{minipage}{0.45\textwidth}
\begin{flushleft}
\hspace{1cm}
\includegraphics[scale=.6]{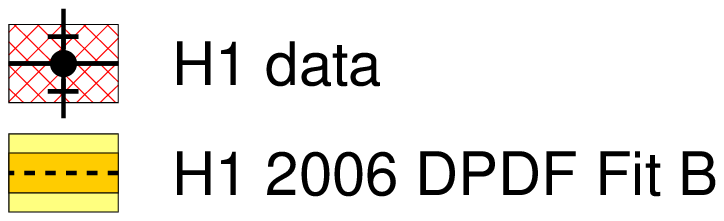}
\end{flushleft}
\includegraphics[width=\textwidth]{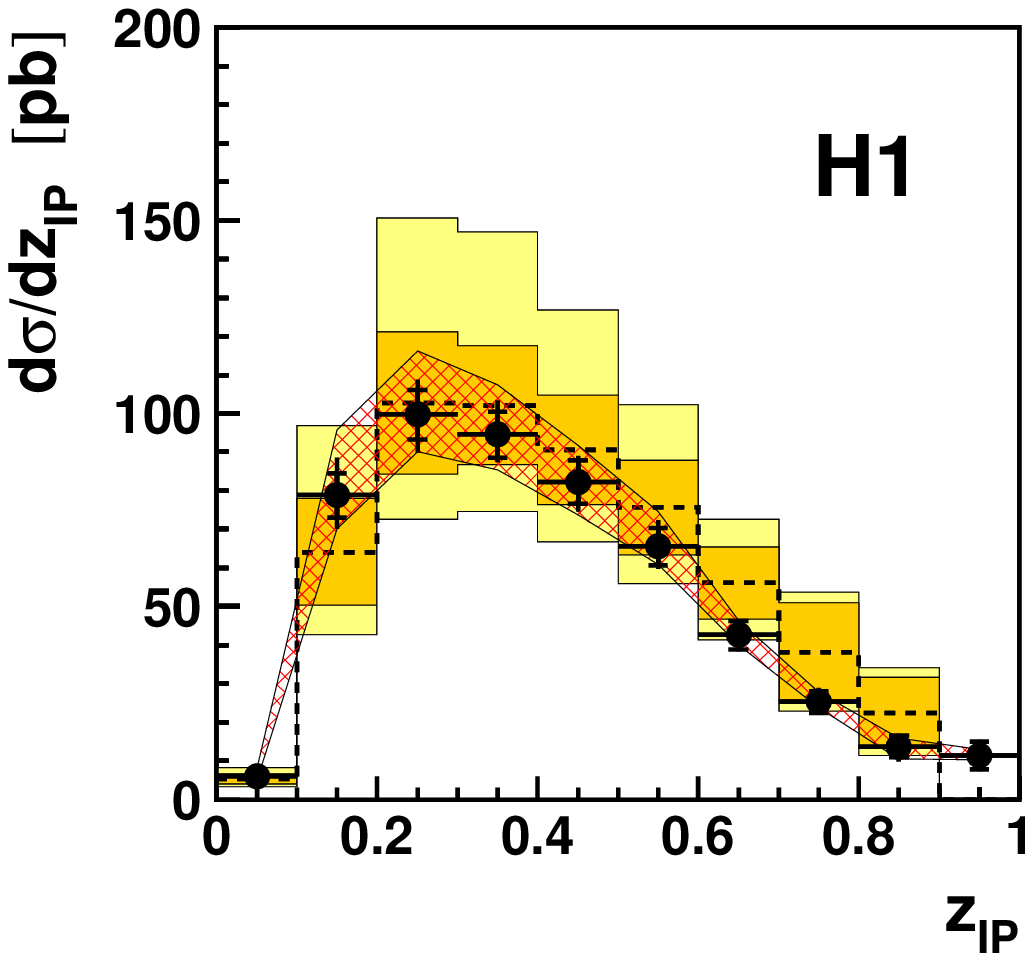}
\end{minipage}
\caption{Cross section for diffractive dijets, differential in $\zpom$
  compared to NLO predictions based on the parton-densities from the
  H1 2006 DPDF fits~\cite{h1f2d97}. The data are shown as black points
  with the inner and outer error bars denoting the statistical and
  quadratically added uncorrelated systematic uncertainties,
  respectively. The hatched band indicates the correlated systematic
  uncertainty. In the left panel the data are compared to the NLO QCD prediction based on
  the H1 2006 DPDF fit A (dotted line) and in the right panerl to the
  prediction based on the  H1 2006 DPDF fit B (dashed line). The lines 
  are surrounded by a dark shaded band indicating the parton density and
  hadronisation uncertainties. In the light shaded band the scale uncertainty is added
  quadratically to the parton density and hadronisation uncertainties. The prediction for $\zpom>0.9$  is not shown since  the hadronisation corrections for this bin cannot be determined reliably.}
\label{fig:zp_fp}
\end{center}
\end{figure}

\begin{figure}[h!]
\hspace{1cm}
\includegraphics[scale=.6]{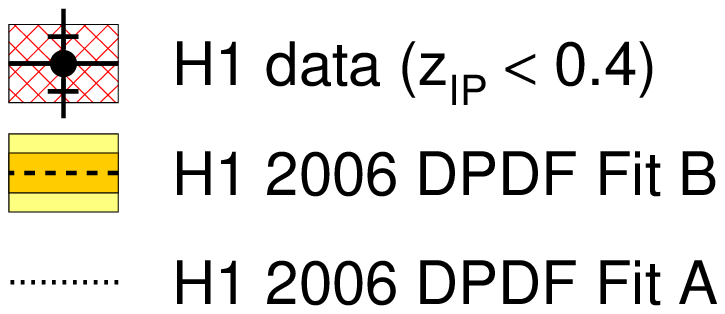}
\begin{center}
\includegraphics[width=.45\textwidth]{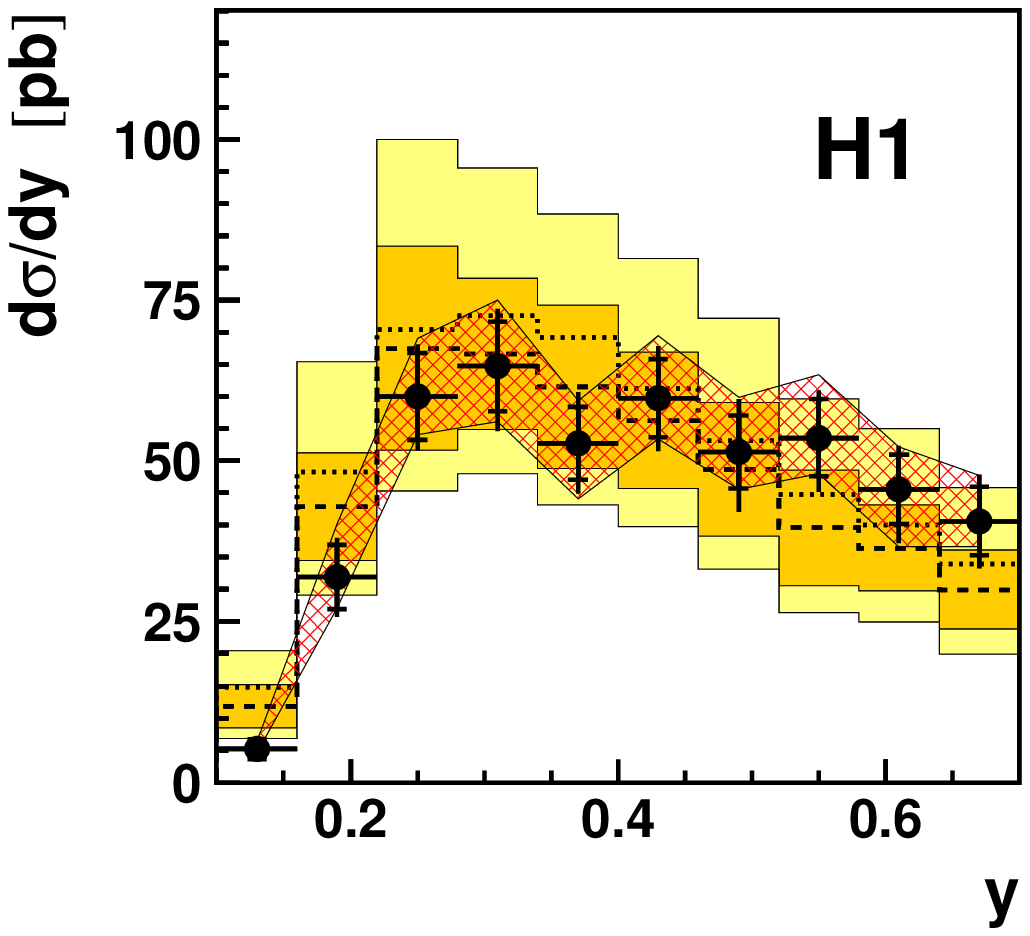}
\hspace{.05\textwidth}
\includegraphics[width=.45\textwidth]{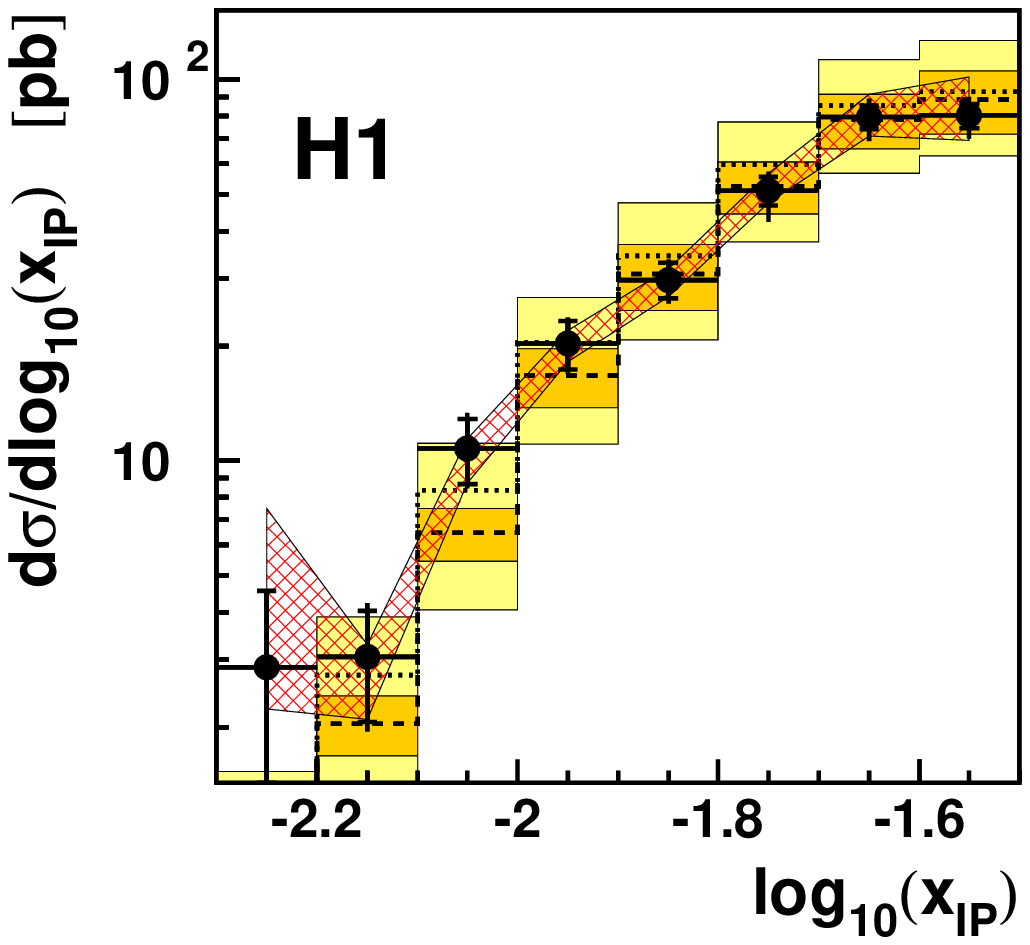}
\vspace{.05\textwidth}\\
\includegraphics[width=.45\textwidth]{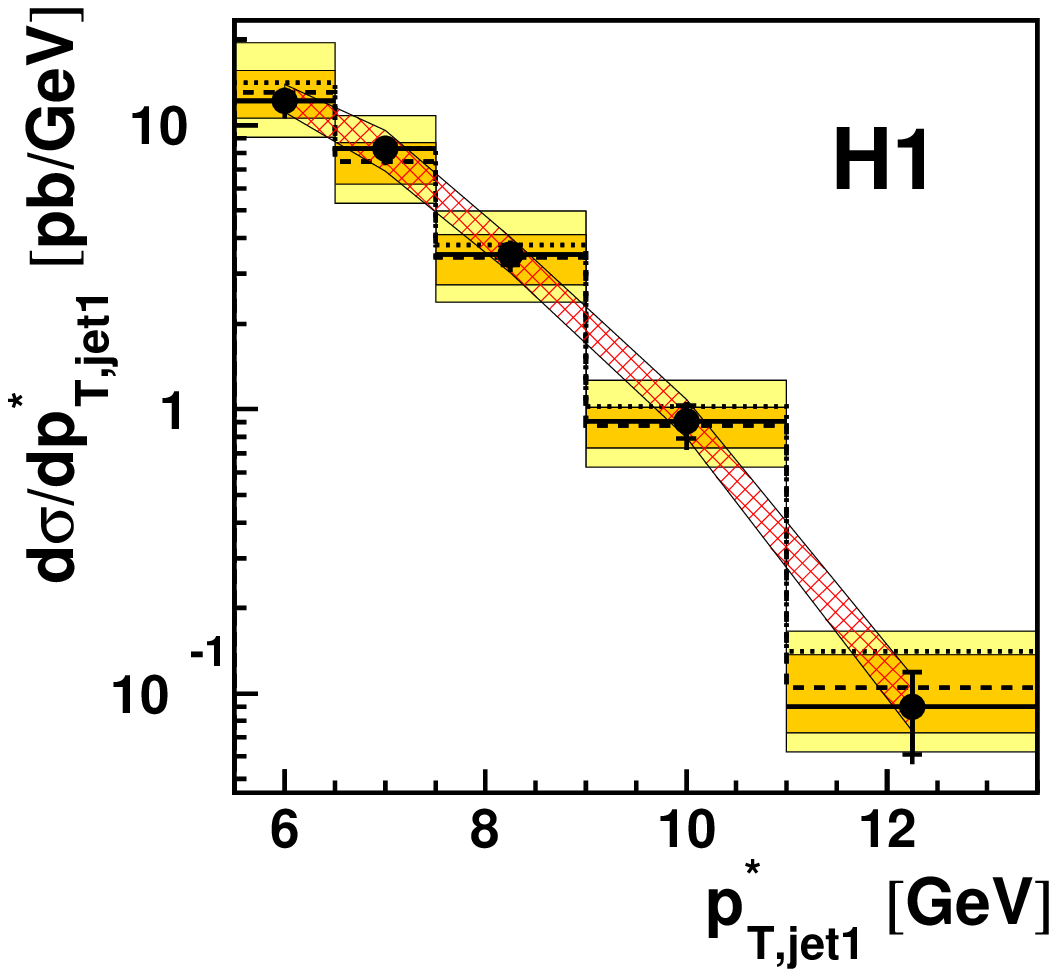}
\hspace{.05\textwidth}
\includegraphics[width=.45\textwidth]{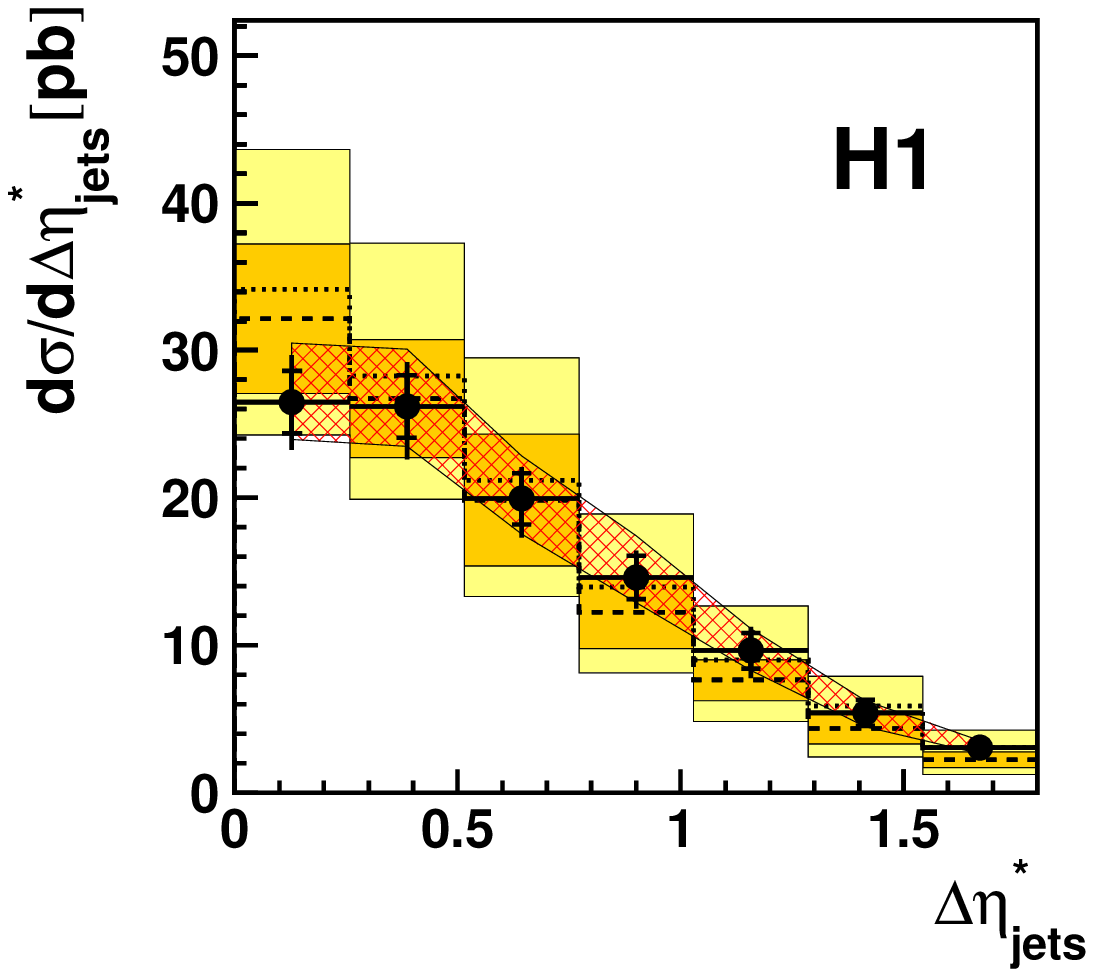}
\caption{Cross sections for diffractive dijets restricted to $\zpom<0.4$, differential in
$y$, $\log\xpom$, $p^\star_{T,jet1}$ and $\Delta\eta^{\star}_{jets}$ compared to NLO
  predictions based on the parton-densities from the H1 2006 DPDF
  fits~\cite{h1f2d97}. The data are shown as black points with the
  inner and outer error bars denoting the statistical and quadratically added uncorrelated
  systematic uncertainties, respectively. The hatched band
  indicates the correlated systematic uncertainty. The dashed
  line shows the NLO QCD prediction based on the H1 2006 DPDF fit B, which is
  surrounded by a dark shaded band indicating the parton density and
  hadronisation uncertainties. In the light shaded band the scale uncertainty is added
  quadratically to the parton density and hadronisation uncertainties.
  The
  dotted line represents the NLO QCD prediction based
  on the H1 2006 DPDF fit A.}
\label{fig:lowzp}
\end{center}
\end{figure}

Differential dijet cross sections are shown in figures~\ref{fig:kin_fp}
to~\ref{fig:eta1} and tabulated in tables~\ref{data_xp} to~\ref{data2}.
Cross sections as a function of $y$, $\xpom$,
$p_{T,jet1}^\star$ and $\Delta\eta_{jets} ^\star = |\eta^\star_{jet1} - \eta^\star_{jet2}|$  are  shown in
Figure~\ref{fig:kin_fp} and are compared to NLO QCD predictions. % assuming QCD factorisation and using the DPDFs extracted in~\cite{h1f2d97}.
Differential cross sections as a function of $\zpom$ are shown in
Figure~\ref{fig:zp_fp}.  %The inner error 
 The NLO QCD
prediction for the highest bin in $\zpom$ is not shown due to problems
in evaluating the
hadronisation corrections\footnote{In some cases the Lund string fragmentation
  algorithm turns the entire system $X$ into just two mesons. This
  leads to events having $\zpom\simeq 1$ at the hadron level
  independently of their parton level $\zpom$ and to corresponding migration problems.}. %Since it is in better agreement with the jet data, fit B is used as the default, with the central result of fit A only  shown for comparison. 

The prediction based on H1 2006 DPDF fit B describes the shapes of all
 distributions well, whereas some discrepancies are apparent between the
 fit A and the data.
 The largest differences between
 the shapes of the two predictions can be seen in $\zpom$ and $y$,
 which are correlated through the kinematics. The discrepancies
 between Fit A and the data are most prominent in the region of high
 $\zpom$ ($\zpom\simgeq 0.4$), where the prediction is clearly too high. 
The good agreement in the $\xpom$ distribution between the dijet data
 and the predictions indicates that the pomeron flux  (which governs
 this distribution) for jet production does not differ significantly from the
 flux describing inclusive diffraction. The shapes of the
 $\Delta\eta_{jets}^\star$ and $p_{T,jet1}^\star$ distributions
 are determined by the hard scattering matrix elements and are rather
 insensitive to the DPDFs. The agreement in these distributions shows
 that the NLO QCD computation, which uses boson gluon fusion as the dominant process,  is adequate to describe dijet production in this kinematic regime.

The large difference between the two predictions at high $\zpom$
   reflects the large uncertainty on the gluon density in this range
  as determined from inclusive data alone. Figure~\ref{fig:zp_fp} also indicates the
  sensitivity of the dijet data  to the gluon density at large $\zpom$.
To test factorisation in a region where the gluon density is well determined from the
inclusive data, the dijet cross section 
is also measured  in the reduced kinematic domain of $\zpom <0.4$.
The results are shown in Figure~\ref{fig:lowzp} and are compared with predictions
based on the H1 2006 DPDF fits. In this kinematic region  both  fits 
   agree well with the dijet data, supporting the notion of
  QCD factorisation within uncertainties.

%%%%%%%%%%%%%%%%%%%%%%%%%%%%%%%%%%%%%%%%%%%%%%%%%%%%%%%%%%%%%%%%%%%%%%%%%%%%%%%%%%%%%

\section{ Combined NLO  QCD Fit} %to the Dijet- and  Inclusive Diffractive  Cross Sections}
A NLO QCD fit is used to determine the diffractive  quark singlet and gluon densities. This combined fit uses both the measurements of the
diffractive dijet cross sections 
presented in this paper and the measurement of the inclusive  diffractive 
cross section presented in~\cite{h1f2d97}. The combined fit
shall henceforth be referred to as `H1 2007 Jets DPDF'.

\subsection{Data Sets}
Assuming the  factorisation hypothesis,
the differential  dijet cross section as a function of $z_{\pom}$
 is used
in the fit in four bins of $Q^2+p^{\star 2}_{T,jet1}$, which is
taken to be the scale variable. These measured cross sections are
shown in figure~\ref{fig:zpbins} and tables~\ref{data1} and~\ref{data2}
for dijets at the stable hadron level in the kinematic range specified in table
\ref{tab:xsdef}.
The fit also includes the measurements of  inclusive diffraction
 obtained by H1 in~\cite{h1f2d97}, which are presented in the form  of the reduced diffractive deep-inelastic scattering cross section $\sigma_r^{D(3)}$, defined through

\begin{equation*}
\frac{d^3\sigma_{ep\rightarrow eXY}}{d\xpom d\beta d\qsq}=\frac{4\pi \alpha_{em}^2}{\beta^2Q^4}\cdot
 Y_+\cdot\sigma_r^{D(3)}(\xpom,\beta,\qsq),
\end{equation*}
where $Y_{+}=1 + (1-y)^2$. In leading order the reduced cross section $\sigma_r^{D(3)}$ is identical to $F_2^{D(3)}$. 
The small influence of the longitudinal structure function
$F_L^{D(3)}$ is included here via its NLO dependence on the DPDFs.
Following the treatment in~\cite{h1f2d97} only data in the range $Q^2
\geq 8.5~\gevsq$, $M_X \geq 2~\gev$  and $\beta \leq 0.8$ are included
in the fit. Figures~\ref{fig:f2d1}~and~\ref{fig:f2d4} show the inclusive
data points  in the form of the product
$\xpom\cdot\sigma_{r}^{D(3)}(\xpom,\beta,\qsq)$.
\subsection{Fit Ansatz}
\label{fit_ansatz}
The DPDFs $f_i^D(z,\mu_f^2,\xpom,t)$ are parameterised following the
fit procedure of the inclusive analysis~\cite{h1f2d97}. They are factorised into a
pomeron flux $f_{\pom /p}(\xpom,t)$ and parton densities of the
pomeron $f_i(z,\mu_f^2)$ using the proton vertex factorisation ansatz

\begin{equation*}
 f_i^D(z,\mu_f^2,\xpom,t) = f_{\pom /p}(\xpom,t) \cdot f_i(z,\mu_f^2).
\end{equation*}

The parton densities $f_i$ are modelled as a singlet distribution $\Sigma(z,\mu_f^2)$ consisting of the three light quark and corresponding antiquark distributions, which are all assumed to be of equal magnitude, and a gluon distribution $g(z,\mu_f^2)$. Here $z$ is the longitudinal momentum fraction of the parton entering the hard subprocess with respect to the diffractive exchange, such that $z=\beta=x/\xpom$ and $z=\zpom$ for the lowest order quark parton model process in inclusive diffraction and for dijets, respectively. The parton densities  $f_i(z,\mu_f^2)$ are parameterised at a starting scale of $\mu_{f,0}^2=2.5~\gevsq$ and are evolved to higher factorisation scales using a numerical solution of the NLO DGLAP evolution equations. 
The singlet and gluon distributions are parameterised at the starting scale as

\begin{equation*}
f_i (z,\mu_{f,0}^2) \equiv A_i \cdot z^{B_i} \cdot (1 - z )^{C_i}.
\end{equation*}
The parameterisation of the singlet density is thus identical to that
used in the analysis of inclusive diffraction~\cite{h1f2d97}. The
parameterisation of the gluon density differs  in that the  H1 2006
DPDF fit A omits the factor $z^{B_{gluon}}$, while fit B omits both $z^{B_{gluon}}$ and $(1-z)^{C_{gluon}}$. %The additional inclusion of $B_{gluon}$ in the combined fit was chosen to allow an extra degree of freedom for the gluon density, which is much more constrained due to the inclusion of the dijet data.
In the H1 2007 Jets DPDF fit, where the dijet data additionally constrain the
gluon, the $\chi^2$ of the fit is significantly reduced by the
inclusion of the factor $z^{B_{gluon}}$.

The pomeron flux  is parameterised as in~\cite{h1f2d97} using a form motivated  by
Regge theory:% with fixed parameters chosen to be identical to those in~\cite{h1f2d97}:

\begin{equation*}
f_{\pom /p}(\xpom,t)=A_{\pom}\left(\frac{1}{\xpom} \right)^{2\alpha_{\pom}(t)-1}e^{B_{\pom}t}.
\end{equation*}
The normalisation parameter $A_{\pom}$ is defined as in~\cite{h1f2d97}. The pomeron trajectory $\alpha_{\pom}(t)$ is assumed to be linear:

\begin{equation*}
\alpha_{\pom}(t)=\alpha_\pom(0)+\alpha'_\pom\cdot t.
\end{equation*}
For comparison with the data, all DPDFs are integrated over the
 measured range  $|t|<1~\gevsq$. % to the minimal value kinematically possible.
 To properly  describe the data, especially at high $\xpom$, it is necessary to include a sub-leading exchange
 (the so called reggeon, $\reg$, for details see~\cite{h1f2d97}). This contribution is assumed to factorise similarly to the pomeron, so that the definition of the diffractive parton densities is modified to

\begin{equation*}
 f_i^D(z,\mu_f^2,\xpom,t) = f_{\pom /p}(\xpom,t) \cdot f_i(z,\mu_f^2) + n_\reg \cdot f_{\reg / p}(\xpom,t) \cdot f_i^{\reg}(z,\mu_f^2).
\end{equation*}
The reggeon flux $f_{\reg / p}(\xpom,t)$ is parameterised in the same
way as the pomeron flux. The parton densities $f_i^{\reg}(z,\qsq)$ are taken from a parameterisation of pion structure function data~\cite{Martin:2002aw}.
The free parameters of the fit are the six parameters of the initial parton densities, $\alpha_\pom(0)$ and the normalisation of the reggeon flux $n_\reg$.
%A summary of the  fixed input parameters is given in table \ref{fitinput}. 
All other parameters are fixed using the same values and uncertainties
as in~\cite{h1f2d97} as listed in table~\ref{fitinput}.
\begin{table}
\begin{center}
\begin{tabular}{|l|r@{}@{}l@{}@{}l|c|}
\hline
\bf Parameter &\multicolumn{3}{c|}{\bf Value} & \bf Source\\
\hline
& & & & \\
$\alpha'_\pom$  & 0.06 & $^{+0.19}_{-0.06}$ &$~\mathrm{GeV^{-2}}$ & \cite{Kapichine}\\
$B_\pom$        & 5.5  & $^{+0.7}_{-2.0} $ &$~\mathrm{GeV^{-2}}$ & \cite{Kapichine}\\
$\alpha_{\reg}(0)$& 0.5  & $\pm   0.1$  &  &   \cite{h1f2d94}\\
$\alpha'_\reg$  & 0.3  & $^{+0.6}_{-0.3}  $ &$~\mathrm{GeV^{-2}}$ & \cite{Kapichine}\\
$B_\reg$        & 1.6  & $^{+0.4}_{-1.6}  $ &$~\mathrm{GeV^{-2}}$ & \cite{Kapichine}\\
$m_c$           & 1.4  & $\pm $  0.2  &~\gev  & \cite{PDG} \\
$m_b$           & 4.5  & $\pm $  0.5  &~\gev  & \cite{PDG}\\
$\alpha_s(M_Z^2)$  & 0.118& $\pm $  0.002&  & \cite{PDG} \\
&&&&\\
\hline
\end{tabular}
\caption{Fixed parameters and associated uncertainties  used in the
  H1 2007 Jets DPDF.}
\label{fitinput}
\end{center}
\end{table}

\subsection{Fit Procedure}
The fit is performed by minimisation of a $\chi^2$ function, defined
similarly to that in~\cite{h1f2d97}. At each step of the minimisation
procedure, the predictions for $\sigma_r^{D(3)}$ are calculated at NLO
in the $\overline{MS}$ renormalisation scheme with the QCDFIT program~\cite{F2evolution,pascaud2}.
For the prediction of the dijet cross section the combined fit uses
the `matrix method' introduced by the ZEUS
collaboration~\cite{ZEUSfit}, together with  the NLOJET++ program.
% The diffractive parton densities are also used to predict  differential dijet cross sections using the NLOJET++ program. 
%This second step would be very time consuming if it were repeated for every fit step. The NLOJET++ predictions are therefore calculated only once for a  large range of values in the variables $\zpom$, $\xpom$ and bins in the scale
% $p_{T,jet1}^{\star 2} + Q^2$ for all  parton flavours and are stored in matrix form using reasonably good starting values for the diffractive
% parton  distributions from previous fits. The predictions for the cross section for each fit step   are then obtained by weighting the matrix elements with the ratio of new to old DPDFs  for each bin.
% All contributions are then summed up to predict the differential jet cross section. If the hard scale is changed  for the fit this can be corrected for by  weighting the matrix with the ratio of new and old values of $\alpha _s$.
This procedure has been shown to yield results which agree with direct NLOJET++
predictions in the selected fit range to better than 2\%  after one iteration of
the input DPDFs (for details see~\cite{mmthesis}). Whereas the
NLOJET++ calculation employs a  massless heavy flavour scheme,  the prediction for
$\sigma_r^D$ is performed with  massive charm and beauty quarks. However, for the dijet
data the hard scale is typically much larger than the charm mass
($\mu^2_{r,f}>29~\gevsq\gg m_c^2$), so  little effect is
expected.  This is confirmed by performing fits to the inclusive data
alone in both schemes, resulting in very similar gluon
densities in the $\qsq$ range to which the dijets are sensitive.

The inclusive and dijet data sets are statistically independent and
 the correlations between the two measurements through the systematic
 uncertainties are small. The $\chi^2$ function treats the combined
 statistical und uncorrelated systematic errors for each data point in
 the usual way and also takes  account of correlated uncertainties
 within the inclusive or dijet data sets by allowing variations in the
 corresponding systematic error sources at the expense of increases in
 the $\chi^2$ variable~\cite{Adloff:2000qk}. As in~\cite{h1f2d97}, there are ten
 such error sources for inclusive data. The correlated errors on the
 jet cross sections are treated via a single additional parameter in
 the $\chi^2$ finction

Besides the uncertainties related to the cross section measurements, the extracted DPDFs are affected
by uncertainties in the fit procedure and its theory input.  The fit
errors include the rather small effects of the uncertainties on the
input parameters as given in table \ref{fitinput}. 
 The uncertainty in the relative scale choice between the
inclusive and dijet data is estimated by  varying the
scale for the dijet data between $2\cdot\sqrt{\qsq+p^{\star
    2}_{T,jet1}}$ and $0.5\cdot\sqrt{\qsq+p^{\star
    2}_{T,jet1}}$ whilst keeping $Q$ as the scale for the inclusive
data.
In addition the effects of changing  the fit range in
 $\zpom$ (excluding dijet events with $\zpom<0.2$) or the starting scale
$\mu_{f,0}^2$ (using $3.5~\gevsq$ instead of $2.5~\gevsq$) are
 evaluated and included in the presented uncertainties. To assess the dependence of the final fitted parton
 densities on the hadronisation correction applied to the dijets, 
 alternative correction factors  extracted from the POMWIG Monte
 Carlo model are used. The deviation from  the
 nominal fit result is included in the theoretical parton density uncertainties.
The largest theoretical error contribution to the fitted gluon density at high $\zpom$ comes
from  the  uncertainty in the relative scale of the two data
 sets. %, the fit-range in $\zpom$ and the hadronisation corrections. 

\section{H1 2007 Jets DPDF Fit Results}
 The fit results for the free parameters are summarized in table \ref{fitresults}.
 The fit describes the data well as indicated by the overall value of
 $\chi^2 / ndf$ = 196/218, which splits into $\chi^2$ = 27 for the 36
 dijet data points and $\chi^2$ =  169 for the 190 $\sigma_r^{D(3)}$ data points.  %Thus the total $\chi^2$ for the inclusive part is 10 units larger in the combined fit than in the fit to inclusive data alone from~\cite{h1f2d97}, indicating a small remaining tension between the two data sets. 
Thus the partial $\chi^2$ for the inclusive data is slightly larger in the combined
 fit than in the fits to $\sigma_r^{D(3)}$ from~\cite{h1f2d97}, where
 $\chi^2$ = 158 (164) for the H1 2006 DPDF fit A (B),  indicating a small remaining tension between the two data sets.
%than the results from the fits to $\sigma_r^{D(3)}$. where $\chi^2$ = 158 (164) for H1 2006 DPDF fit A (B),  indicating a small remaining tension between the two data sets. 
The parameter $C_{gluon}$, determining the gluon density behaviour at high values
of $z$, is  positive in the combined fit in accordance
with the expectation that the gluon density should not be singular for
$z\rightarrow 1$. This behaviour is different from the H1 DPDF fit A,
where $C_{gluon}$ is determined to be negative and the gluon density
 is artificially suppressed at very  high $z$ using an additional exponential factor.

\begin{table}[h!]
\begin{center}
\begin{tabular}{|l|l@{}@{}c@{}@{}l|}
\hline
\bf Parameter &\multicolumn{3}{c|} {\bf Fit Value} \\
 &\multicolumn{3}{c|} {\bf (H1 2007 Jets DPDF)} \\
\hline
$\alpha_{\pom}(0)$ & 1.104 & $\pm$ & 0.007\\
$n_\reg$    & 1.3$\times 10^{-3}$ & $\pm$ & $0.4\times 10^{-3}$\\
$A_{gluon}$ & 0.88 & $\pm$ & 0.17\\
$B_{gluon}$ & 0.33 & $\pm$ & 0.10\\
$C_{gluon}$ & 0.91 & $\pm$ & 0.18\\
$A_{quark}$ & 0.13 & $\pm$ & 0.02\\
$B_{quark}$ & 1.5  & $\pm$ & 0.12\\
$C_{quark}$ & 0.51 & $\pm$ & 0.08\\
\hline
$\chi^2/ndf$      & 196/218 & &\\
\hline
\end{tabular}
\caption{H1 2007 Jets DPDF fit parameters obtained from the combined fit to the diffractive inclusive and dijet data. Only the experimental uncertainties are given.}
\label{fitresults}
\end{center}
\end{table}

\begin{figure}[h!]
\hspace{2cm}
\includegraphics[scale=0.6]{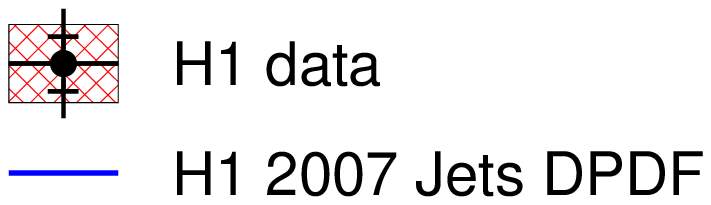}
\begin{center}
\includegraphics[width=\textwidth]{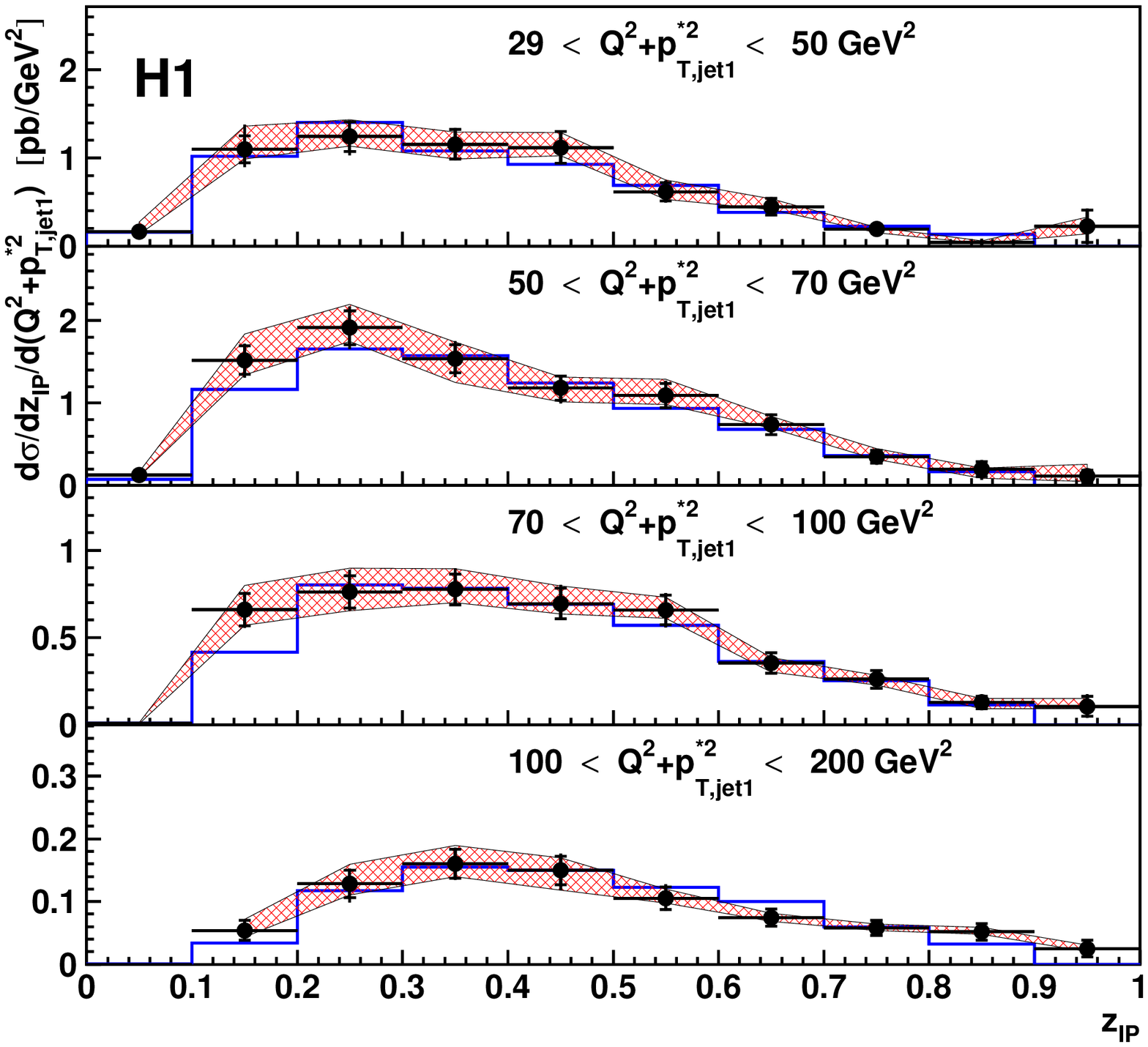}
\caption{Cross section for diffractive dijet production doubly differential in
  $\zpom$ and the scale $\qsq+p_{T,jet1}^{\star 2}$. The data are shown
  as black points with the inner and outer error bars denoting the
  statistical and quadratically added  uncorrelated systematic uncertainties,
  respectively. The hatched band indicates the correlated systematic
  uncertainty. The solid line shows the NLO QCD prediction based on the
  H1 2007 Jets DPDF. Data points in the highest $\zpom$ bin were not
  included in the fit since  the hadronisation corrections cannot be evaluated reliably.}
\label{fig:zpbins}
\end{center}
\end{figure}

\begin{figure}[h!]
\hspace{1cm}
\includegraphics[scale=0.6]{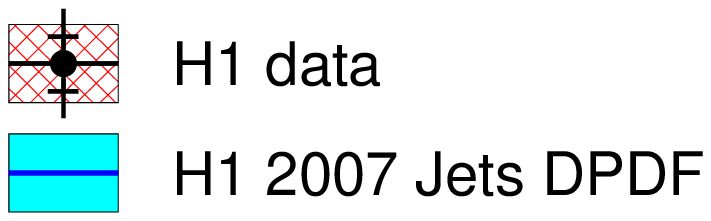}
\begin{center}
\includegraphics[width=.45\textwidth]{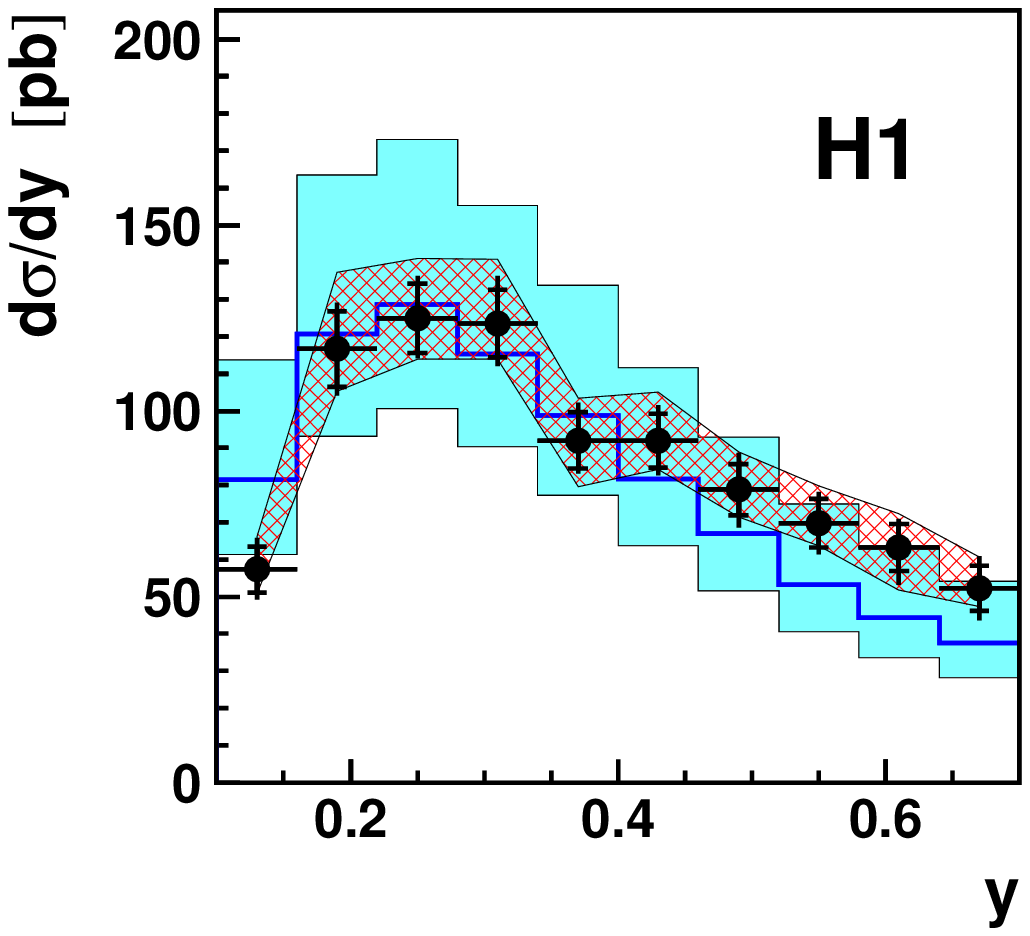}
\hspace{.05\textwidth}
\includegraphics[width=.45\textwidth]{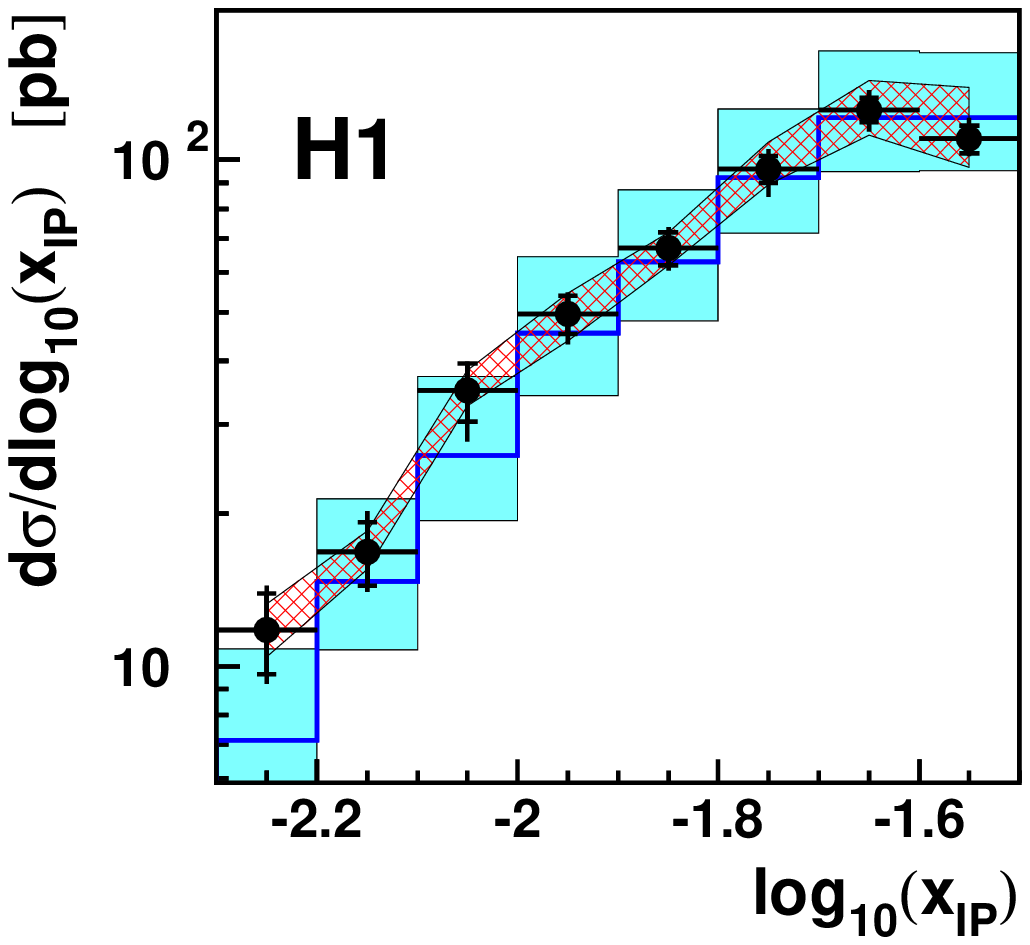}
\vspace{.05\textwidth}\\
\includegraphics[width=.45\textwidth]{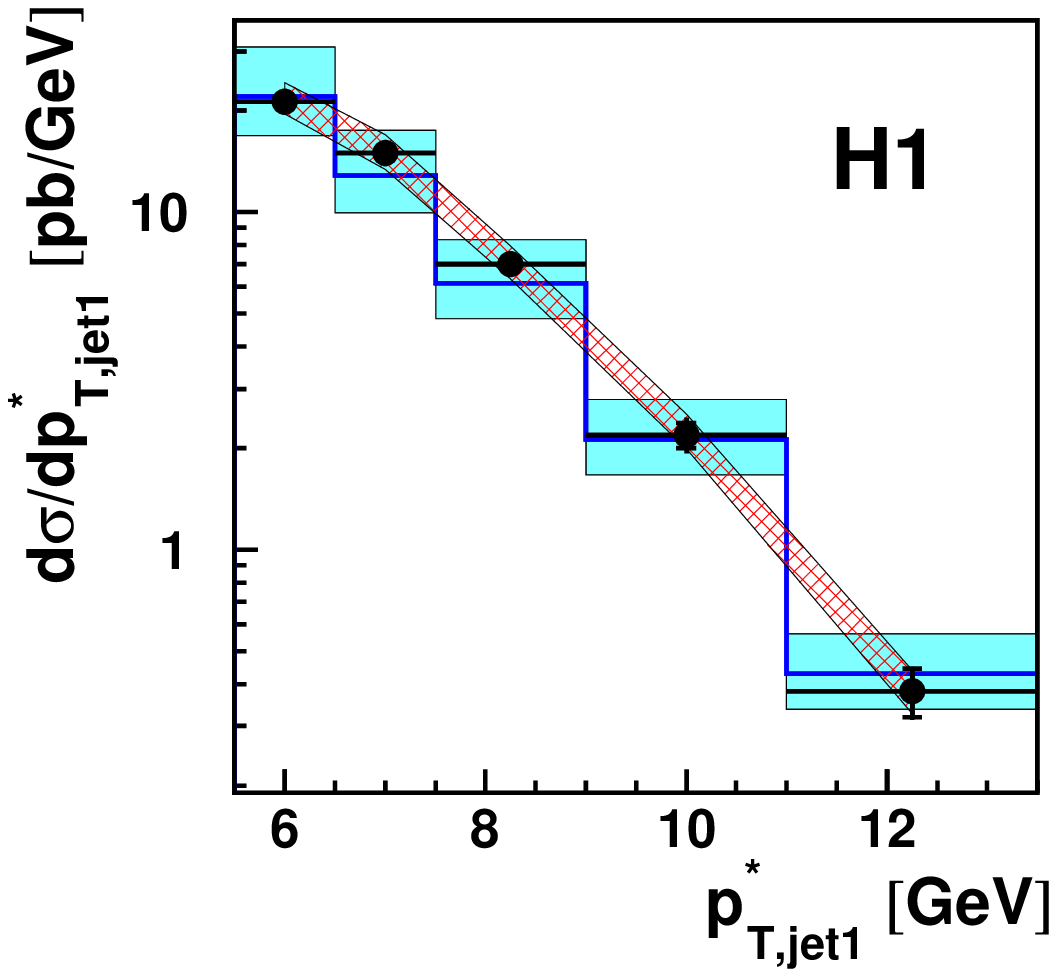}
\hspace{.05\textwidth}
\includegraphics[width=.45\textwidth]{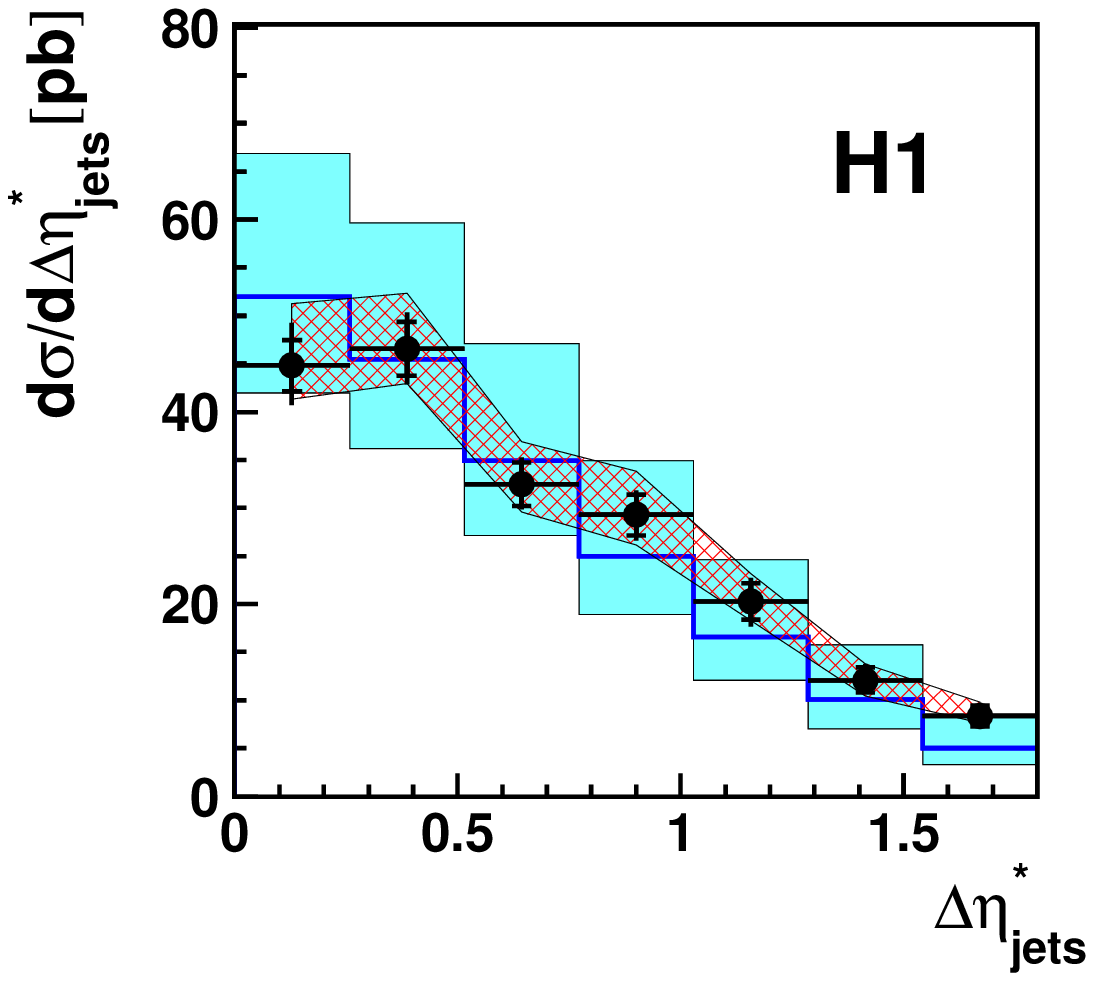}
\caption{Cross sections for diffractive dijet production differential in the variables $y$, $\log\xpom$, $p_{T,jet1}^\star$ and $\Delta\eta^\star_{jets}$. The data are shown as black points with the inner and outer error bars denoting the statistical and quadratically added uncorrelated systematic uncertainties, respectively. The hatched band indicates the correlated systematic uncertainty. The solid line surrounded by the shaded band shows the NLO QCD prediction based on the H1 2007 Jets DPDF, where the band denotes the scale uncertainty derived by varying the renormalisation and factorisation scale $\mu=\sqrt{\qsq+p^2_{T,jet1}}$ by factors of 2 and 0.5.}
\label{fig:eta1}
\end{center}
\end{figure}

\begin{sidewaysfigure}[h!]
\begin{center}
\includegraphics[width=0.45\textwidth]{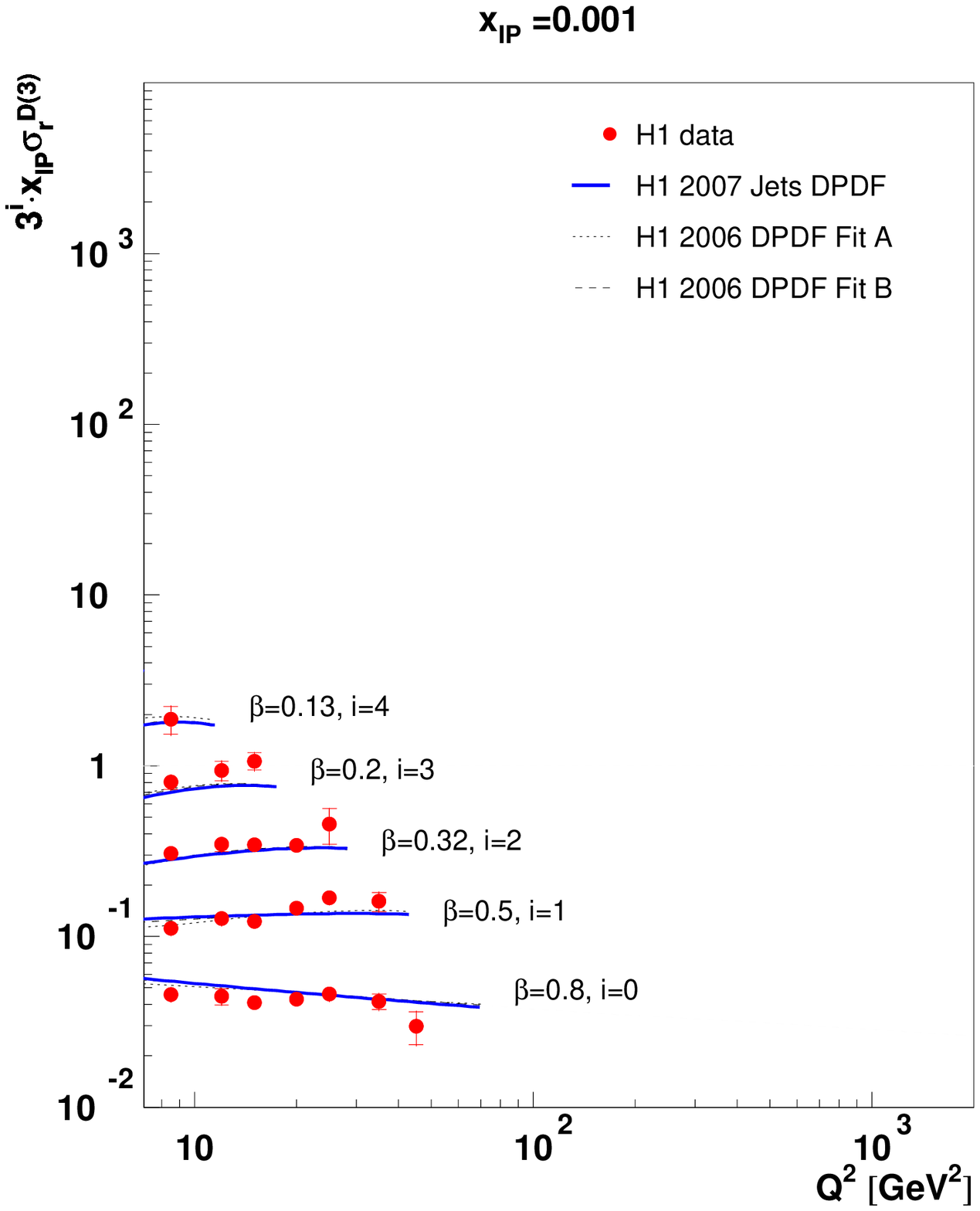}
\hspace{2cm}
\includegraphics[width=0.45\textwidth]{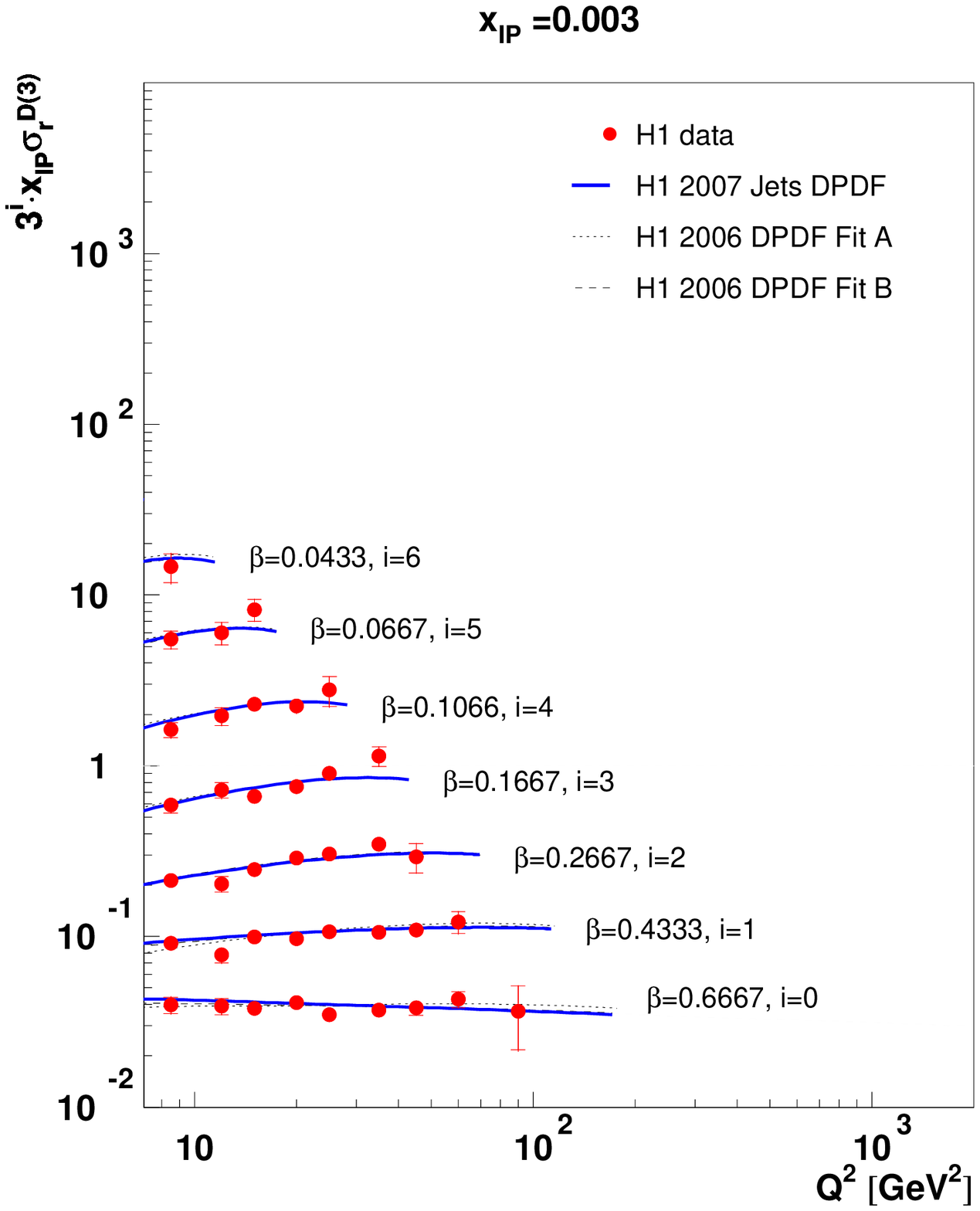}
\caption{The $\qsq$ dependence of the diffractive reduced
  cross section $\sigma_r^{D(3)}$ multiplied by \xpom \ at
  \xpom=0.001 (left) and \xpom=0.003 (right) at various values of $\beta$. The cross sections are multiplied by powers of 3 for
  better visibility. The data points are taken from the publication~\cite{h1f2d97}. The inner and outer error bars on the data points
  represent the statistical and total uncertainties,
  respectively. Only data points included in the DPDF fits are shown. The
  data are compared to NLO QCD predictions based on the H1 2007 Jets DPDF, which are shown as solid lines. The dashed and dotted lines indicate the
  predictions of the H1 2006 DPDF fit A and B, respectively.}
\label{fig:f2d1}
\end{center}
\end{sidewaysfigure}

\begin{sidewaysfigure}[h!]
\begin{center}
\includegraphics[width=0.45\textwidth]{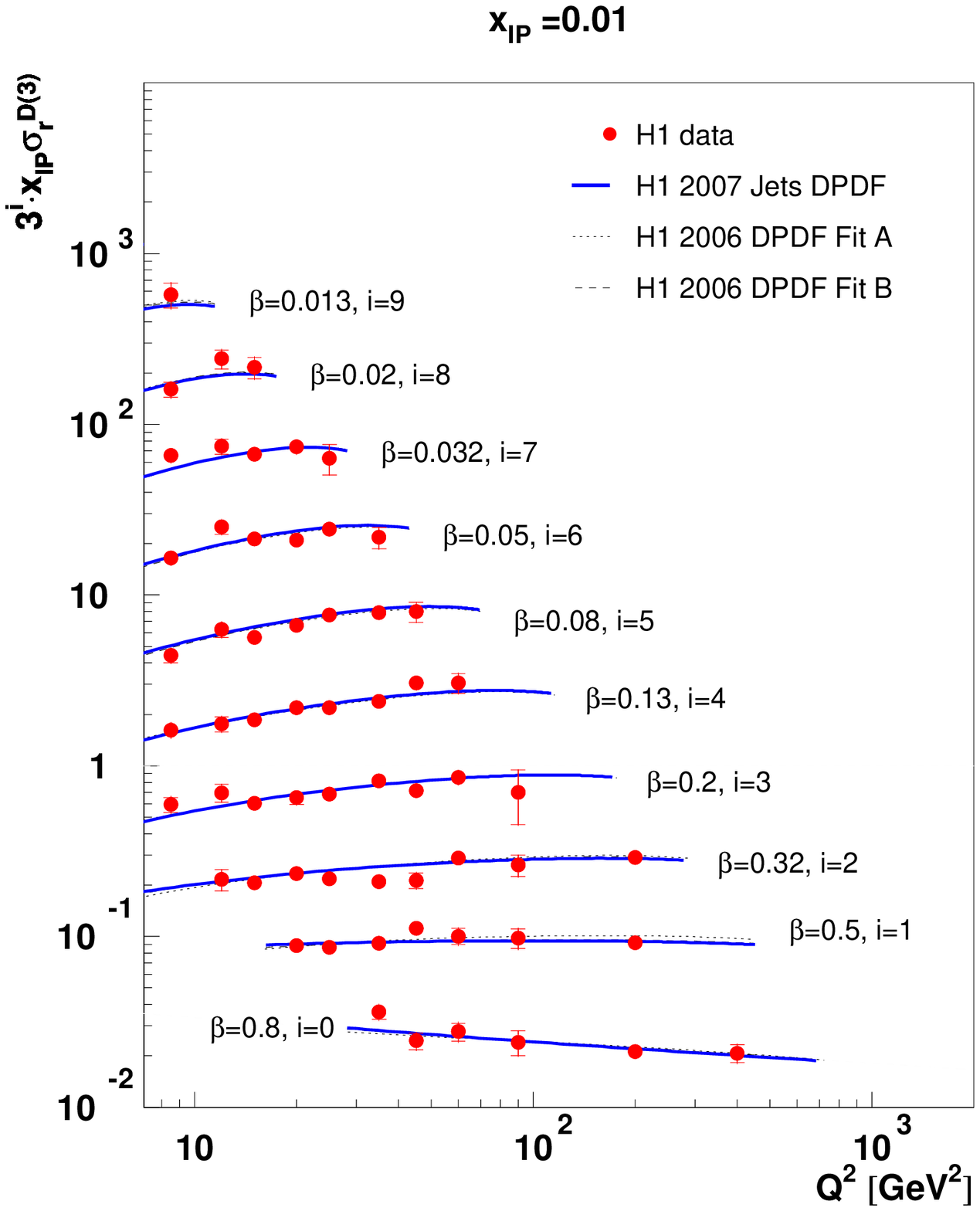}
\hspace{2cm}
\includegraphics[width=0.45\textwidth]{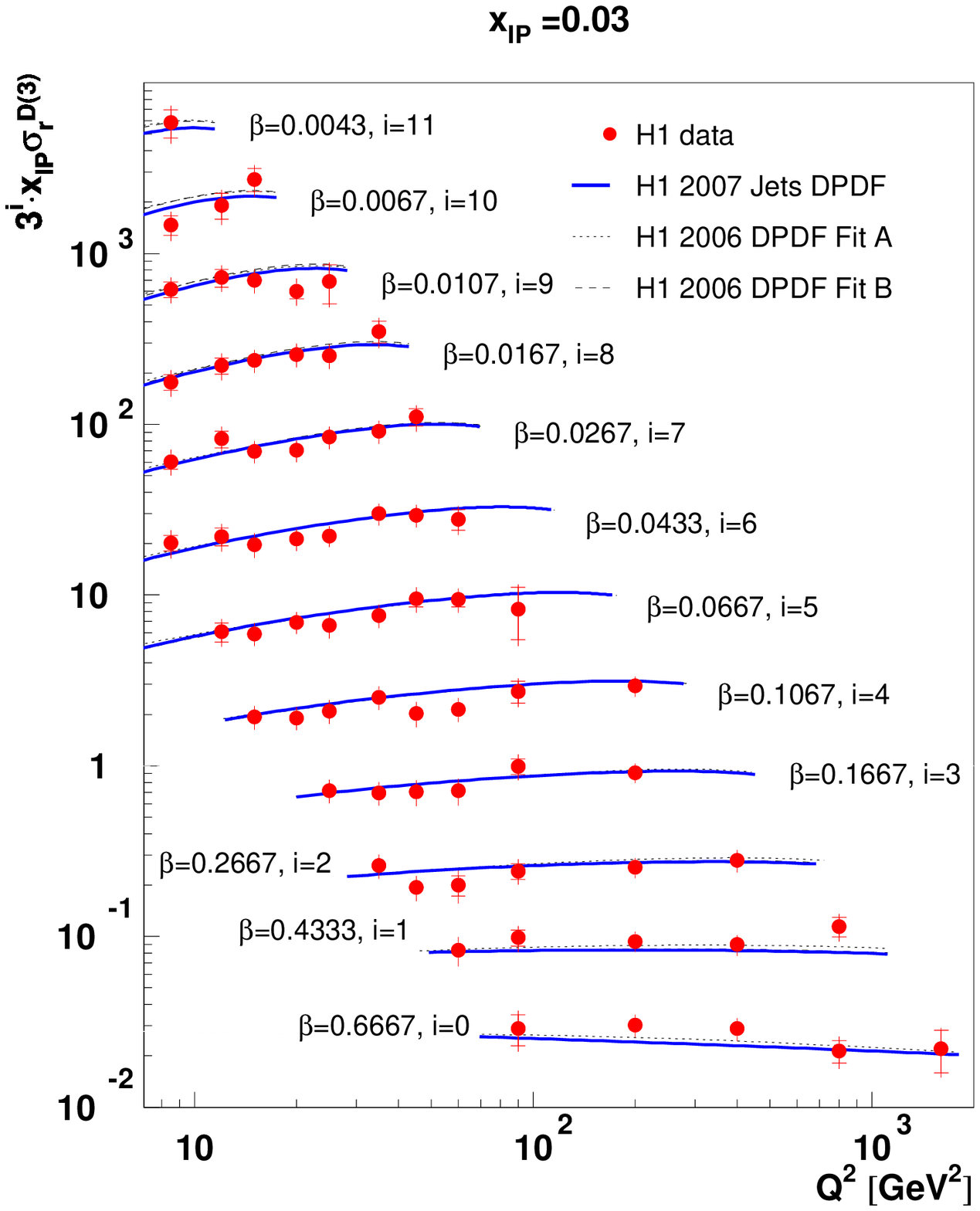}
\caption{The $\qsq$ dependence of the diffractive reduced cross section $\sigma_r^{D(3)}$ multiplied by \xpom \ at \xpom=0.01 (left) and \xpom=0.03 (right) at various values of $\beta$. See caption of figure~\ref{fig:f2d1} for further details.}
\label{fig:f2d4}
\end{center}
\end{sidewaysfigure}

The dijet cross sections are well described by the predictions based
on the H1 2007 Jets DPDF as shown in figures \ref{fig:zpbins}
and \ref{fig:eta1}. %The error bands on the NLO predictions are scale errors obtained by varying the
Figures~\ref{fig:f2d1} and~\ref{fig:f2d4} show the measurements of
 $\sigma_r^{D(3)}$ as a function of  $\qsq$ for different
 values of $\beta$ and  $x_{\pom}$, together with the NLO  predictions based on the
 H1 2007 Jets DPDF fit. The results of fits A
 and B to the  inclusive data alone are also shown.
A very good description is obtained with
 all three fits. % which is also reflected in the resulting values of $\chi ^2 / ndf$  given above. 

\begin{figure}[h!]
\hspace{1cm}
\includegraphics[scale=0.6]{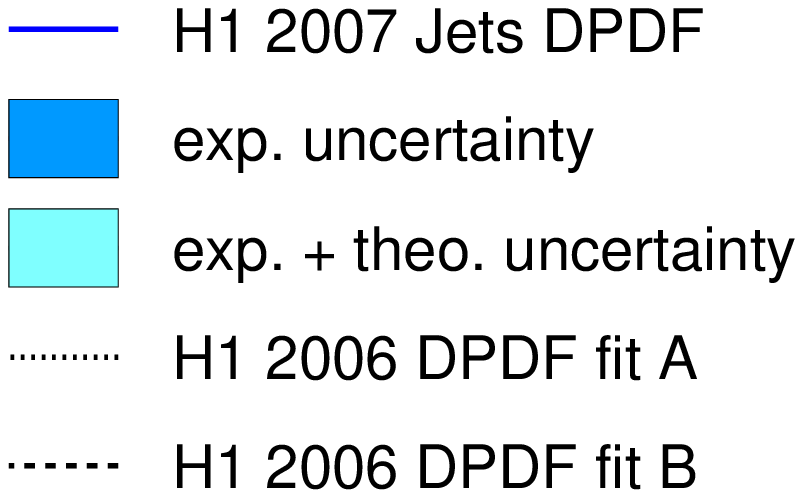}
\begin{center}
\includegraphics[width=.45\textwidth]{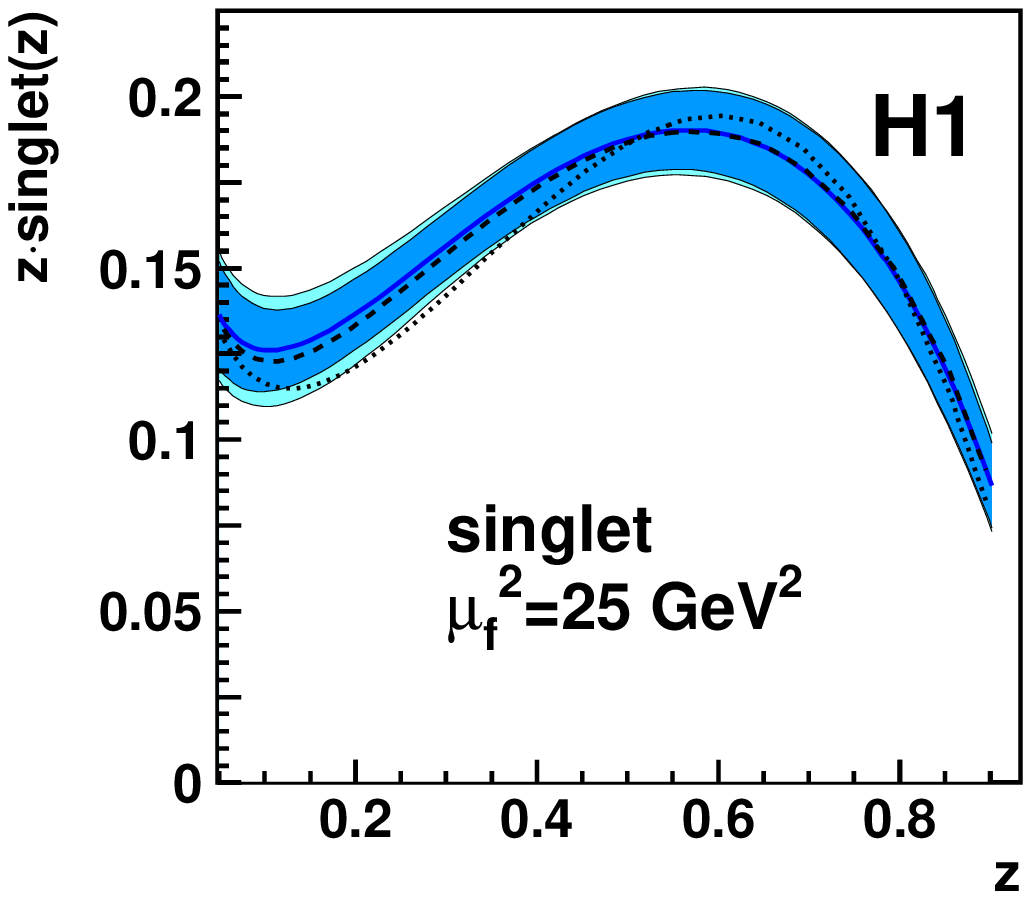}
\hspace{.05\textwidth}
\includegraphics[width=.45\textwidth]{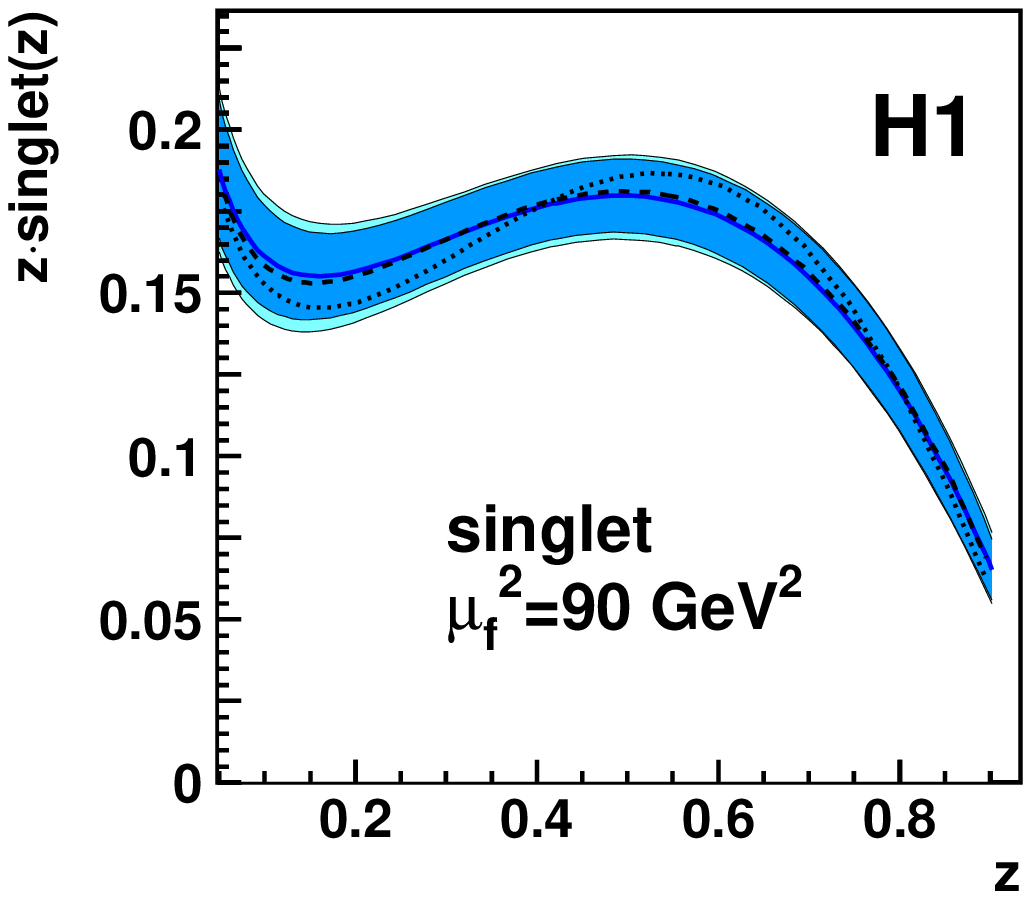}
\vspace{.05\textwidth}\\
\includegraphics[width=.45\textwidth]{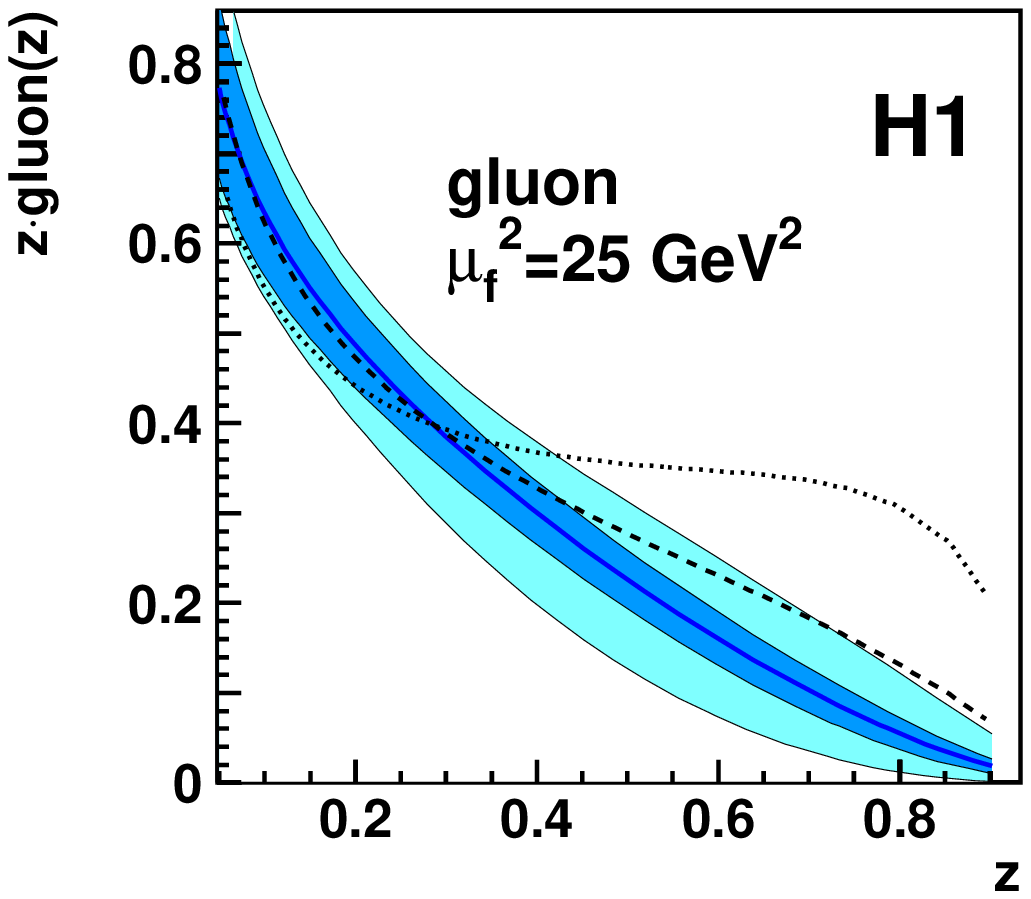}
\hspace{.05\textwidth}
\includegraphics[width=.45\textwidth]{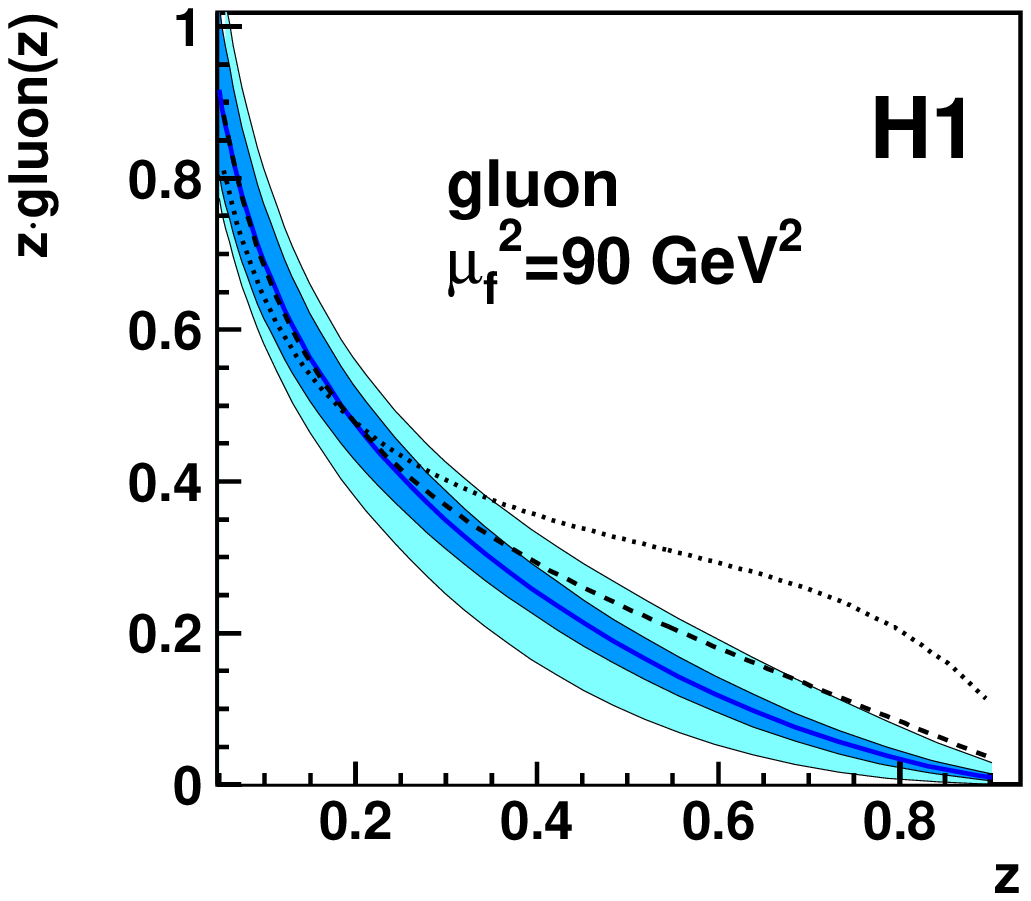}
\caption{The diffractive quark density (top) and the diffractive
  gluon density (bottom) for two values of the squared factorisation scale
  $\mu_f^2$: 25~\gevsq (left) and 90~\gevsq (right). The solid line
  indicates the H1 2007 Jets DPDF, surrounded by the experimental
  uncertainty (dark shaded band) and the experimental and theoretical
  uncertainties added in quadrature (light shaded band). The dotted and dashed lines show the parton densities corresponding to 
the H1 2006 fit A and fit B  from~\cite{h1f2d97}, respectively.  }
\label{fig:gluon}
\end{center}
\end{figure}

 The diffractive gluon distribution and the
 quark singlet distribution are shown in figure \ref{fig:gluon} for
  scales of $\mu_f ^2 = 25~\gevsq$ and $\mu_f ^2 = 90~\gevsq$, together with the
 results of fits A and B of the stand-alone analysis of
 $\sigma_r^{D(3)}$. The error bands indicate the uncertainties due to
 experimental sources and
 the theoretical errors inherent in the fit procedure.
  The  uncertainties on the quark distribution and on the gluon distribution at low $\zpom$ are dominated by the experimental uncertainties, while the uncertainty on the gluon density at high $\zpom$ recieves sizeable contributions from both experimental and theoretical sources.

The combined fit constrains both the diffractive gluon and quark densities 
 well and for the first time with comparable precision in the complete range $0.05 < z_{\pom} < 0.9$. 
At high $\zpom$ the resulting gluon density differs significantly from
that of
H1 2006 DPDF fit A, but is compatible with fit B~\cite{h1f2d97}. Good
agreement is seen between all three fits for the singlet quark density
and the gluon density at low $\zpom$. 
The values of $\alpha_{\pom}(0)= 1.104 \pm  0.007$ and  $n_\reg =
(1.3\pm0.4)\times 10^{-3}$  are  compatible
within experimental uncertainties with the value extracted in H1 2006
fit B. The uncertainties on these parameters are not significantly
decreased by the inclusion of the dijet data compared to the
determination from the inclusive data alone.

\begin{figure}[h!]
\hspace{1cm}
\includegraphics[scale=0.6]{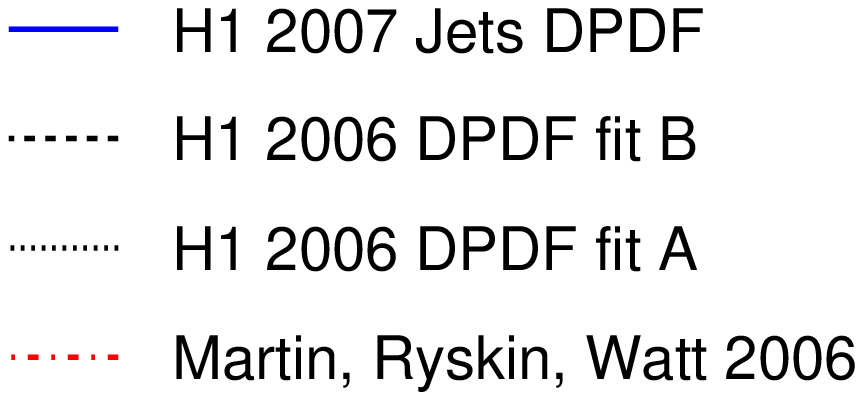}
\begin{center}
\includegraphics[width=.45\textwidth]{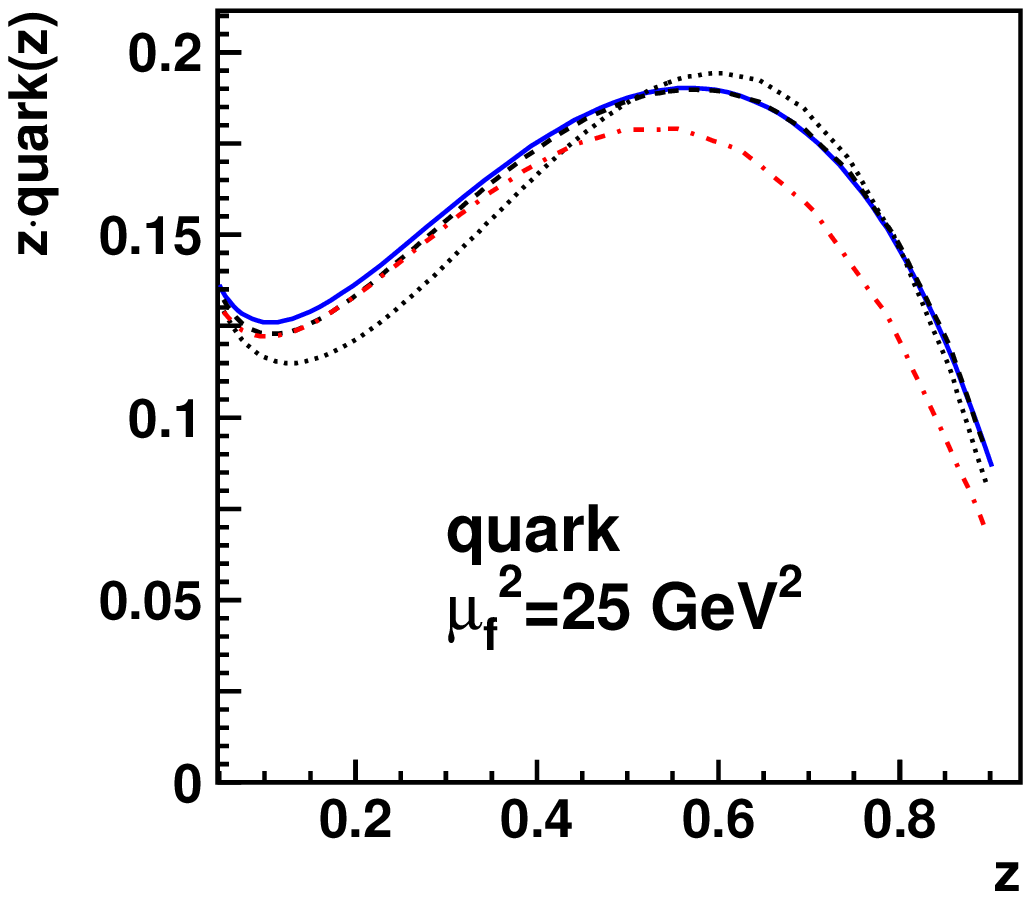}
\hspace{.05\textwidth}
\includegraphics[width=.45\textwidth]{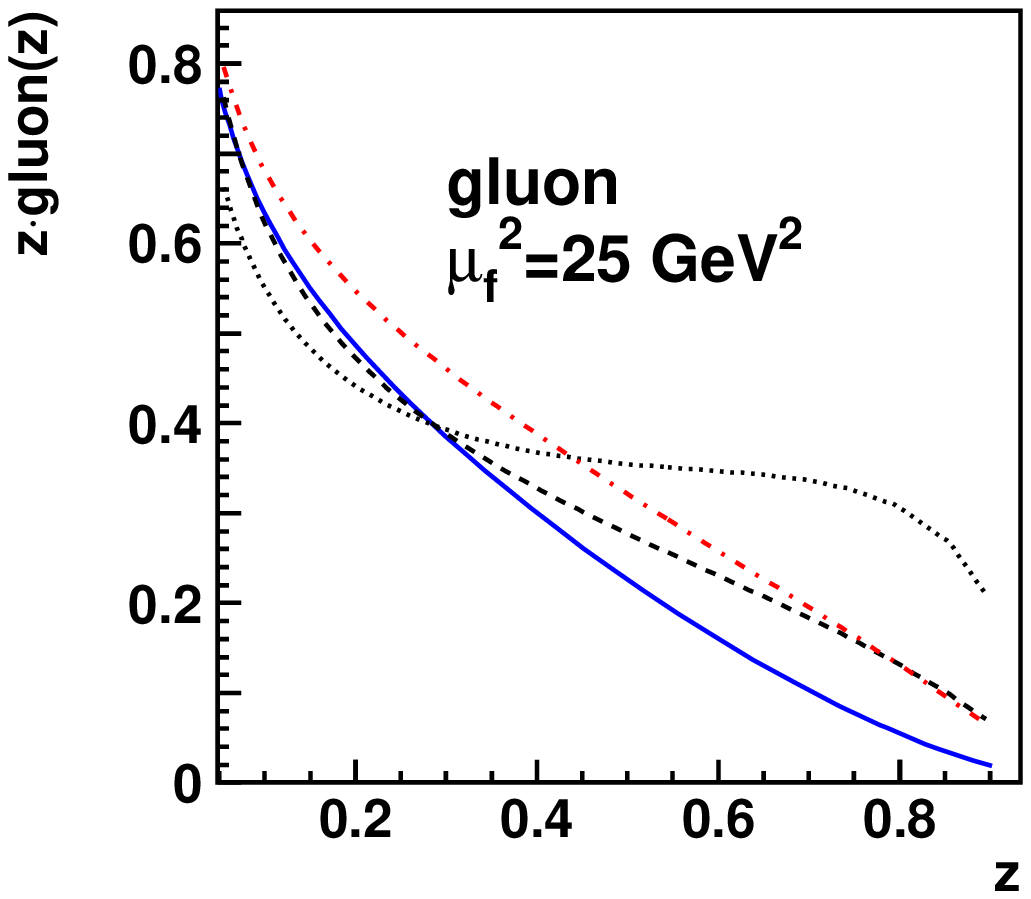}
\caption{The diffractive quark density (left) and the diffractive  gluon density (right) for the factorisation scale $\mu_f^2=25$~\gevsq. The solid line indicates the H1 2007 Jets DPDF. The dotted and dashed lines show the parton densities corresponding to 
the H1 2006 fit A and fit B  from~\cite{h1f2d97}, respectively. The
dashed-dotted line shows the DPDFs as determined by Martin, Ryskin and
Watt in~\cite{Martin:2006td}.}
\label{fig:mrw}
\end{center}
\end{figure}

In figure~\ref{fig:mrw} the
DPDFs as determined by H1 are compared with the results of an
independent analysis~\cite{Martin:2006td}, where 
parton densities are derived from the same inclusive diffractive data~\cite{h1f2d97}. A hybrid theoretical framework is used which combines
aspects of collinear factorisation and a perturbative
two-gluon-exchange model~\cite{bjklw,bjklw_2,bjklw_3,bjklw_4}. Most of
the dijets events are produced via  BGF-type processes as in figure~\ref{feynman}. At high $\beta$, there is an additional
contribution in which the perturbative
two-gluon state participates directly in the hard interaction
via photon-pomeron fusion,
leading to a modified evolution equation for the DPDFs. The
resulting DPDFs agree reasonably well with the H1 2007 Jets DPDF  and
with the 
H1 2006 DPDF fit B.% but is notably different from H1 DPDF fit A.

Measurements of diffractive charm production by H1 have also been
compared to predictions based on the DPDFs presented in this
paper~\cite{temp}. Whilst overall good agreement is obtained, 
the statistical accuracy of the charm measurement limits its power to
discriminate between different DPDF sets.

\section{Conclusion}

Cross sections for dijet production in diffractive deep-inelastic
scattering are  measured with improved precision
compared to earlier analyses. Single and double differential cross sections are presented  in a variety of variables  sensitive to the underlying dynamics of hard diffraction.
NLO QCD predictions based on diffractive
parton densities extracted from measurements of inclusive diffractive  deep-inelastic scattering
describe the data
well in the kinematic region where the gluon density is reliably
constrained by the inclusive measurements.
This agreement confirms the
validity of QCD factorisation  and thus the applicability of diffractive parton
 densities evolving according to the DGLAP equations.

A combined fit to diffractive inclusive and dijet data is
performed, using NLO QCD calculations based on QCD factorisation and DGLAP evolution. Both data sets
are described well by the fit.  
The inclusion of the dijet data allows %for the first time 
the simultaneous determination of  both the diffractive gluon and  the singlet quark distribution with  good and comparable
 accuracy in the range $0.05 < z_{\pom} < 0.9$.  This is the first reliable
 determination of the diffractive gluon density up to large momentum fractions.

%%%%%%%%%%%%%%%%%%%%%%%%%%%%%%%%%%%%%%%%%%%%%%%%%%%%%%%%%%%%%%%%%%%%%%%%%%%%%%%%%%%%%
%\input{acknowledgements}
%%%%%%%%%%%%%%%%%%%%%%%%%%%%%%%%%%%%%%%%%%%%%%%%%%%%%%%%%%%%%%%%%%%%%%%%%%%%%%%%%%%%%
\section*{Acknowledgments}
We are grateful to the HERA machine group whose outstanding
efforts have made this experiment possible.
We thank
the engineers and technicians for their work in constructing and 
maintaining the H1 detector, our funding agencies for
financial support, the
DESY technical staff for continual assistance,
and the DESY directorate for the support and the
hospitality which they extend to the non DESY
members of the collaboration.

%%%%%%%%%%%%%%%%%%%%%%%%%%%%%%%%%%%%%%%%%%%%%%%%%%%%%%%%%%%%%%%%%%%%%%%%%%%%%%%%%%%%%
%\input{bibliography}
%%%%%%%%%%%%%%%%%%%%%%%%%%%%%%%%%%%%%%%%%%%%%%%%%%%%%%%%%%%%%%%%%%%%%%%%%%%%%%%%%%%%%

\clearpage
%%%%%%%%%%%%%%%%%%%%%%%%%%%%%%%%%%%%%%%%%%%%%%%%%%%%%%%%%%%%%%%%%%%%%%%%%%%%%%%%%%%%%
%\input{figures}
%%%%%%%%%%%%%%%%%%%%%%%%%%%%%%%%%%%%%%%%%%%%%%%%%%%%%%%%%%%%%%%%%%%%%%%%%%%%%%%%%%%%%
%%%%%%%%%%%%%%%%%%%%%%%%%%%%%%%%%%%%%%%%%%%%%%%%%%%%%%%%%%%%%%%%%%%%%%%%%%%%%%%%%%%%%
%%%%%%%%%%%%%%%%%%%%%%%%%%%%%%%%%%%%%%%%%%%%%%%%%%%%%%%%%%%%%%%%%%%%%%%%%%%%%%%%%%%%%
%%%%%%%%%%%%%%%%%%%%%%%%%%%%%%%%%%%%%%%%%%%%%%%%%%%%%%%%%%%%%%%%%%%%%%%%%%%%%%%%%%%%%%%
%%%%%%%%%%%%%%%%%%%%%%%%%%%%%%%%%%%%%%%%%%%%%%%%%%%%%%%%%%%%%%%%%%%%%%%%%%%%%%%%%%%%%%%
%%%%%%%%%%%%%%%%%%%%%%%%%%%%%%%%%%%%%%%%%%%%%%%%%%%%%%%%%%%%%%%%%%%%%%%%%%%%%%%%%%%%%%%
%%%%%%%%%%%%%%%%%%%%%%%%%%%%%%%%%%%%%%%%%%%%%%%%%%%%%%%%%%%%%%%%%%%%%%%%%%%%%%%%%%%%%%%
%%%%%%%%%%%%%%%%%%%%%%%%%%%%%%%%%%%%%%%%%%%%%%%%%%%%%%%%%%%%%%%%%%%%%%%%%%%%%%%%%%%%%%%
%%%%%%%%%%%%%%%%%%%%%%%%%%%%%%%%%%%%%%%%%%%%%%%%%%%%%%%%%%%%%%%%%%%%%%%%%%%%%%%%%%%%%%%
%%%%%%%%%%%%%%%%%%%%%%%%%%%%%%%%%%%%%%%%%%%%%%%%%%%%%%%%%%%%%%%%%%%%%%%%%%%%%%%%%%%%%%%
%%%%%%%%%%%%%%%%%%%%%%%%%%%%%%%%%%%%%%%%%%%%%%%%%%%%%%%%%%%%%%%%%%%%%%%%%%%%%%%%%%%%%%%

\clearpage

\setlength{\tabcolsep}{3mm}
\begin{table}[h!]
\footnotesize
\begin{center}
\begin{tabular}{|r@{\hspace{3mm}}l@{\hspace{3mm}}l|c|c|c|c|c|r@{\hspace{3mm}}c@{\hspace{3mm}}l|}
\hline
\multicolumn{3}{|c|}{$\log(x_\pom)$} & $d\sigma/d\log(\xpom)$ & $\delta_{\mathrm{tot.}}$&$\delta_{\mathrm{stat.}}$ & $\delta_{\mathrm{uncorr.}}$  &
$\delta_{\mathrm{corr.}}$  & \multicolumn{3}{c|}{hadr.}\\
\multicolumn{3}{|c|}{} &[pb] &[pb] &[pb] &[pb] &[pb] &\multicolumn{3}{c|}{ corr.} \\
\hline
-2.3 & - & -2.2 & 11.8& 3.0& 2.1&  1.5& 1.4&1.55 &$\pm$& 0.15\\
-2.2 & - & -2.1 & 16.8& 3.4& 2.4&  1.9& 1.5&1.38 &$\pm$& 0.02\\
-2.1 & - & -2.0 & 35.0& 6.5& 4.6&  3.6& 2.9&1.24 &$\pm$& 0.02\\
-2.0 & - & -1.9 & 49.6& 7.8& 4.3&  3.7& 5.3&1.24 &$\pm$& 0.04\\
-1.9 & - & -1.8 & 66.8& 8.2& 4.9&  4.2& 4.9&1.10 &$\pm$& 0.06\\
-1.8 & - & -1.7 & 96  & 14 & 6  &  8  & 9  &1.11 &$\pm$& 0.04\\
-1.7 & - & -1.6 & 125 & 20 & 7  &  10 & 16 &1.04 &$\pm$& 0.13\\
-1.6 & - & -1.5 & 110 & 23 & 7  &  6  & 21 &1.04 &$\pm$& 0.06\\
\hline
\end{tabular}
\caption{Bin averaged differential cross sections of diffractive dijet production  at the hadron level
  (corrected to the  QED Born level) and the corresponding uncertainties as a function of $\xpom$.  The corrections applied to the NLO
  prediction for hadronisation and the associated uncertainty are also
  given.}
\label{data_xp}
\end{center}
\end{table}

\begin{table}[h!]
\footnotesize
\begin{center}
\begin{tabular}{|r@{\hspace{3mm}}l@{\hspace{3mm}}l|c|c|c|c|c|r@{\hspace{3mm}}c@{\hspace{3mm}}l|}
\hline
\multicolumn{3}{|c|}{$y$} & $d\sigma/dy$  & $\delta_{\mathrm{tot.}}$  & $\delta_{\mathrm{stat.}}$  & $\delta_{\mathrm{uncorr.}}$  &
$\delta_{\mathrm{corr.}}$ & \multicolumn{3}{c|}{hadr.}\\
\multicolumn{3}{|c|}{}  &[pb] &[pb] &[pb] &[pb] &[pb] &\multicolumn{3}{c|}{ corr.}\\
\hline
0.1  & - & 0.16 & 57  & 11 & 6  & 5  & 8  &1.16 &$\pm$& 0.14\\  
0.16 & - & 0.22 & 117 & 20 & 10 & 7  & 16 &1.09 &$\pm$& 0.03\\
0.22 & - & 0.28 & 125 & 18 & 9  & 6  & 13 &1.10 &$\pm$& 0.02\\
0.28 & - & 0.34 & 123 & 18 & 9  & 8  & 14 &1.09 &$\pm$& 0.07\\
0.34 & - & 0.40 & 92  & 15 & 8  & 6  & 12 &1.10 &$\pm$& 0.10\\
0.40 & - & 0.46 & 92  & 14 & 7  & 6  & 10 &1.12 &$\pm$& 0.01\\
0.46 & - & 0.52 & 79  & 13 & 7  & 7  & 9  &1.13 &$\pm$& 0.15\\
0.52 & - & 0.58 & 70  & 12 & 7  & 5  & 8  &1.11 &$\pm$& 0.14\\
0.58 & - & 0.64 & 63  & 14 & 6  & 6  & 10 &1.11 &$\pm$& 0.12\\
0.64 & - & 0.7  & 52  & 11 & 6  & 6  & 7  &1.11 &$\pm$& 0.10\\
\hline
\end{tabular}
\caption{Bin averaged differential cross sections of diffractive dijet production  at the hadron level
  (corrected to the  QED Born level) and the corresponding uncertainties as a function of $y$.  The corrections applied to the NLO
  prediction for hadronisation and the associated uncertainty are also
  given.}
\label{data_y}
\end{center}
\end{table}

\begin{table}[h!]
\footnotesize
\begin{center}
\begin{tabular}{|r@{\hspace{3mm}}l@{\hspace{3mm}}l|c|c|c|c|c|r@{\hspace{3mm}}c@{\hspace{3mm}}l|}
\hline
\multicolumn{3}{|c|}{$\zpom$} & $d\sigma/d\zpom$  & $\delta_{\mathrm{tot.}}$& $\delta_{\mathrm{stat.}}$ & $\delta_{\mathrm{uncorr.}}$  &
$\delta_{\mathrm{corr.}}$ & \multicolumn{3}{c|}{hadr.}\\
\multicolumn{3}{|c|}{}  &[pb] &[pb] &[pb] &[pb] &[pb] &\multicolumn{3}{c|}{ corr.} \\
\hline
0.0 & - & 0.1 & 6.0  & 2.2 & 1.4 & 0.7 & 1.6 &1.28 &$\pm$& 0.18 \\  
0.1 & - & 0.2 & 79   & 16  & 6   & 7   & 13  &1.09 &$\pm$& 0.10 \\
0.2 & - & 0.3 & 100  & 16  & 6   & 7   & 13  &1.10 &$\pm$& 0.06 \\
0.3 & - & 0.4 & 95   & 14  & 6   & 5   & 11  &1.08 &$\pm$& 0.03 \\
0.4 & - & 0.5 & 82   & 12  & 6   & 4   & 9   &1.11 &$\pm$& 0.03 \\
0.5 & - & 0.6 & 65.5 & 9.2 & 4.8 & 4.0 & 6.8 &1.12 &$\pm$& 0.01 \\
0.6 & - & 0.7 & 42.6 & 5.2 & 3.7 & 1.8 & 3.1 &1.09 &$\pm$& 0.09 \\
0.7 & - & 0.8 & 25.3 & 4.0 & 2.8 & 2.0 & 2.0 &0.99 &$\pm$& 0.28 \\
0.8 & - & 0.9 & 13.7 & 4.5 & 2.7 & 2.4 & 2.7 &0.90 &$\pm$& 0.32 \\
0.9 & - & 1.0 & 11.4 & 4.2 & 3.5 & 1.9 & 1.3 &   &--&     \\
\hline
\end{tabular}
\caption{Bin averaged differential cross sections of diffractive dijet production  at the hadron level
  (corrected to the  QED Born level) and the corresponding uncertainties as a function of $\zpom$.  The corrections applied to the NLO
  prediction for hadronisation and the associated uncertainty are also
  given. No hadronisation correction is given for the highest $\zpom$
  bin since it cannot be evaluated reliably.}
\label{data_zp}
\end{center}
\end{table}

\begin{table}[h!]
\footnotesize
\begin{center}
\begin{tabular}{|r@{\hspace{3mm}}l@{\hspace{3mm}}l|c|c|c|c|c|r@{\hspace{3mm}}c@{\hspace{3mm}}l|}
\hline
\multicolumn{3}{|c|}{$p_{T,jet1}^\star$}& $d\sigma/dp_{T,jet1}^\star$& $\delta_{\mathrm{tot.}}$ & $\delta_{\mathrm{stat.}}$ & $\delta_{\mathrm{uncorr.}}$ &
$\delta_{\mathrm{corr.}}$ & \multicolumn{3}{c|}{hadr.}\\
\multicolumn{3}{|c|}{[\gev]}& [pb/\gev] & [pb/\gev] & [pb/\gev] & [pb/\gev] & [pb/\gev]&\multicolumn{3}{c|}{ corr.}\\
\hline
5.5   & - & 6.5  & 21.2 & 3.0  & 1.0  & 1.6   & 2.3   &1.09 &$\pm$& 0.12\\
6.5   & - & 7.5  & 15.0 & 2.2  & 0.8  & 1.0   & 1.8   &1.11 &$\pm$& 0.06\\
7.5   & - & 9.0  & 7.0  & 1.0  & 0.4  & 0.4   & 0.8   &1.11 &$\pm$& 0.01\\
9.0   & - & 11.0 & 2.18 & 0.38 & 0.19 & 0.19  & 0.26  &1.17 &$\pm$& 0.15\\
11.0  & - & 13.5 & 0.38 & 0.088& 0.062& 0.028 & 0.056 &1.12 &$\pm$& 0.08\\
\hline
\end{tabular}
\caption{Bin averaged differential cross sections of diffractive dijet production  at the hadron level
  (corrected to the  QED Born level) and the corresponding uncertainties as a function of $p_{T,jet1}^\star$.  The corrections applied to the NLO
  prediction for hadronisation and the associated uncertainty are also
  given.}
\label{data_pt}
\end{center}
\end{table}

\begin{table}[h!]
\footnotesize
\begin{center}
\begin{tabular}{|r@{\hspace{3mm}}l@{\hspace{3mm}}l|c|c|c|c|c|r@{\hspace{3mm}}c@{\hspace{3mm}}l|}
\hline
\multicolumn{3}{|c|}{$\Delta\eta^\star_{jets}$} & $d\sigma/d\Delta\eta^\star_{jets}$ & $\delta_{\mathrm{tot.}}$ & $\delta_{\mathrm{stat.}}$ &$\delta_{\mathrm{uncorr.}}$  &
$\delta_{\mathrm{corr.}}$ & \multicolumn{3}{c|}{hadr.}\\
\multicolumn{3}{|c|}{} &[pb] &[pb] &[pb] &[pb] &[pb] &\multicolumn{3}{c|}{ corr.}\\
\hline
0     & - & 0.257 & 44.8 & 6.6 & 2.6 & 3.4 & 5.0 &1.05 &$\pm$& 0.05\\  
0.257 & - & 0.514 & 46.6 & 6.0 & 2.8 & 2.4 & 4.7 &1.11 &$\pm$& 0.01\\
0.514 & - & 0.771 & 32.5 & 4.6 & 2.3 & 1.5 & 3.7 &1.12 &$\pm$& 0.04\\
0.771 & - & 1.029 & 29.3 & 4.6 & 2.1 & 1.5 & 3.8 &1.14 &$\pm$& 0.10\\
1.029 & - & 1.286 & 20.3 & 3.5 & 1.9 & 1.6 & 2.4 &1.15 &$\pm$& 0.03\\
1.286 & - & 1.543 & 12.1 & 2.2 & 1.3 & 0.8 & 1.7 &1.20 &$\pm$& 0.13\\
1.543 & - & 1.8   & 8.4  & 1.6 & 1.1 & 0.6 & 1.1 &1.10 &$\pm$& 0.14\\
\hline
\end{tabular}
\caption{Bin averaged differential cross sections of diffractive dijet production  at the hadron level
  (corrected to the  QED Born level) and the corresponding uncertainties as a function of $\Delta\eta^\star_{jets}$.  The corrections applied to the NLO
  prediction for hadronisation and the associated uncertainty are also
  given.}
\label{data_deleta}
\end{center}
\end{table}
\begin{table}[h!]
\footnotesize
\begin{center}
\begin{tabular}{|r@{\hspace{3mm}}c@{\hspace{3mm}}l|c|c|c|c|c|r@{\hspace{3mm}}c@{\hspace{3mm}}l|}
\hline
\multicolumn{11}{|c|}{}\\
\multicolumn{11}{|c|}{\boldmath$29~\gevsq<\qsq+p^2_{T,jet1}<50~\gevsq$}\\
\multicolumn{11}{|c|}{}\\
\hline
\multicolumn{3}{|c|}{$z_\pom$} & \centering{$d^2\sigma/d\zpom d\mu^2$}  & $\delta_{\mathrm{tot.}}$ & $\delta_{\mathrm{stat.}}$ & $\delta_{\mathrm{uncorr.}}$  & $\delta_{\mathrm{corr.}}$  & \multicolumn{3}{c|}{hadr.} \\
& & &[pb/$\gevsq$] & [pb/$\gevsq$]& [pb/$\gevsq$]& [pb/$\gevsq$]& [pb/$\gevsq$]& \multicolumn{3}{c|}{ corr.} \\
\hline
0.0 & - & 0.1 & 0.16  & 0.10  & 0.05  & 0.04 & 0.07  & 1.32   &$\pm$& 0.04\\
0.1 & - & 0.2 & 1.10  & 0.30  & 0.15  & 0.18 & 0.19  & 0.99   &$\pm$& 0.35\\
0.2 & - & 0.3 & 1.24  & 0.25  & 0.17  & 0.10 & 0.15  & 1.09   &$\pm$& 0.11\\ 
0.3 & - & 0.4 & 1.16  & 0.24  & 0.17  & 0.07 & 0.15  & 0.97   &$\pm$& 0.14\\ 
0.4 & - & 0.5 & 1.12  & 0.23  & 0.18  & 0.06 & 0.13  & 1.08   &$\pm$& 0.01\\
0.5 & - & 0.6 & 0.61  & 0.17  & 0.10  & 0.07 & 0.11  & 1.11   &$\pm$& 0.10\\ 
0.6 & - & 0.7 & 0.45  & 0.12  & 0.09  & 0.05 & 0.06  & 0.91   &$\pm$& 0.01\\ 
0.7 & - & 0.8 & 0.197 & 0.071 & 0.056 & 0.031& 0.030 & 0.86   &$\pm$& 0.60\\
0.8 & - & 0.9 & 0.042 & 0.036 & 0.022 & 0.024& 0.015 & 0.98   &$\pm$& 0.50\\
0.9 & - & 1.0 & 0.22  & 0.22  & 0.18  & 0.09 & 0.09  &        &  -- &    \\
\hline
\hline
\multicolumn{11}{|c|}{}\\
\multicolumn{11}{|c|}{\boldmath$50~\gevsq<\qsq+p^2_{T,jet1}<70~\gevsq$}\\[1mm]
\multicolumn{11}{|c|}{}\\
\hline
\multicolumn{3}{|c|}{$z_\pom$} & $d^2\sigma/d\zpom d\mu^2$  & $\delta_{\mathrm{tot.}}$ & $\delta_{\mathrm{stat.}}$ & $\delta_{\mathrm{uncorr.}}$  & $\delta_{\mathrm{corr.}}$  &  \multicolumn{3}{c|}{hadr.} \\
& & & [pb/$\gevsq$] & [pb/$\gevsq$]& [pb/$\gevsq$]& [pb/$\gevsq$]& [pb/$\gevsq$]&\multicolumn{3}{c|}{ corr.} \\
\hline
0.0 & - & 0.1 & 0.124 & 0.059 & 0.047 & 0.018 & 0.030 & 1.21 &$\pm$& 0.70\\
0.1 & - & 0.2 & 1.52  & 0.32  & 0.18  & 0.11  & 0.25  & 1.10 &$\pm$& 0.10\\
0.2 & - & 0.3 & 1.91  & 0.34  & 0.20  & 0.15  & 0.23  & 1.08 &$\pm$& 0.12\\ 
0.3 & - & 0.4 & 1.54  & 0.33  & 0.17  & 0.14  & 0.25  & 1.14 &$\pm$& 0.02\\ 
0.4 & - & 0.5 & 1.18  & 0.23  & 0.15  & 0.09  & 0.15  & 1.07 &$\pm$& 0.20\\
0.5 & - & 0.6 & 1.09  & 0.23  & 0.15  & 0.08  & 0.15  & 1.08 &$\pm$& 0.12\\ 
0.6 & - & 0.7 & 0.74  & 0.15  & 0.12  & 0.05  & 0.07  & 1.13 &$\pm$& 0.08\\
0.7 & - & 0.8 & 0.35  & 0.12  & 0.08  & 0.05  & 0.06  & 0.92 &$\pm$& 0.36\\  
0.8 & - & 0.9 & 0.20  & 0.13  & 0.09  & 0.07  & 0.06  & 0.76 &$\pm$& 0.65\\ 
0.9 & - & 1.0 & 0.11  & 0.13  & 0.07  & 0.02  & 0.10  &     & --   &     \\
\hline

\end{tabular}
\caption{Bin averaged double differential cross sections of diffractive dijet production  at the hadron level
  (corrected to the  QED Born level) and the corresponding uncertainties as a function of $\zpom$ in different bins of $\mu^2=\qsq+p_{T, jet1}^2$.   The corrections applied to the NLO
  prediction for hadronisation and the associated uncertainty are also
  given. No hadronisation correction is given for the highest $\zpom$
  bin since it cannot be evaluated reliably.}
\label{data1}
\end{center}
\end{table}

\begin{table}[h!]
\footnotesize
\begin{center}
\begin{tabular}{|r@{\hspace{3mm}}l@{\hspace{3mm}}l|c|c|c|c|c|r@{\hspace{3mm}}c@{\hspace{3mm}}l|}
\hline
\multicolumn{11}{|c|}{}\\
\multicolumn{11}{|c|}{\boldmath$70~\gevsq<\qsq+p^2_{T,jet1}<100~\gevsq$}\\
\multicolumn{11}{|c|}{}\\
\hline
\multicolumn{3}{|c|}{$z_\pom$} & $d^2\sigma/d\zpom d\mu^2
$  & $\delta_{\mathrm{tot.}}$ &
$\delta_{\mathrm{stat.}}$ & $\delta_{\mathrm{uncorr.}}$  &
$\delta_{\mathrm{corr.}}$  & \multicolumn{3}{c|}{hadr.}  \\
%& & & $/d(\qsq+p^2_{T,jet1})$& & & & & & &   \\
& & & [pb/$\gevsq$] & [pb/$\gevsq$]& [pb/$\gevsq$]& [pb/$\gevsq$]& [pb/$\gevsq$]&  \multicolumn{3}{c|}{ corr.}  \\
\hline
0.0 & - & 0.1 & 0.0096& 0.0083& 0.0069& 0.0036& 0.0028& 1.27 &$\pm$& 0.47 \\
0.1 & - & 0.2 & 0.66  & 0.15  & 0.09  & 0.04  & 0.11  & 1.13 &$\pm$& 0.01 \\
0.2 & - & 0.3 & 0.76  & 0.17  & 0.09  & 0.07  & 0.12  & 1.11 &$\pm$& 0.08 \\ 
0.3 & - & 0.4 & 0.78  & 0.14  & 0.09  & 0.06  & 0.10  & 1.05 &$\pm$& 0.10 \\ 
0.4 & - & 0.5 & 0.69  & 0.13  & 0.09  & 0.04  & 0.081 & 1.12 &$\pm$& 0.05 \\  
0.5 & - & 0.6 & 0.66  & 0.11  & 0.09  & 0.05  & 0.069 & 1.17 &$\pm$& 0.04 \\  
0.6 & - & 0.7 & 0.354 & 0.075 & 0.058 & 0.025 & 0.041 & 1.11 &$\pm$& 0.04 \\
0.7 & - & 0.8 & 0.261 & 0.063 & 0.051 & 0.022 & 0.028 & 1.07 &$\pm$& 0.38 \\
0.8 & - & 0.9 & 0.129 & 0.047 & 0.035 & 0.011 & 0.029 & 0.86 &$\pm$& 0.30 \\
0.9 & - & 1.0 & 0.106 & 0.074 & 0.057 & 0.037 & 0.030 &    &--&     \\
\hline
\hline
\multicolumn{11}{|c|}{}\\
\multicolumn{11}{|c|}{\boldmath$100~\gevsq<\qsq+p^2_{T,jet1}<200~\gevsq$}\\
\multicolumn{11}{|c|}{}\\
\hline
\multicolumn{3}{|c|}{$z_\pom$} & $d^2\sigma/d\zpom d\mu^2$ & $\delta_{\mathrm{tot.}}$ & $\delta_{\mathrm{stat.}}$ & $\delta_{\mathrm{uncorr.}}$  & $\delta_{\mathrm{corr.}}$  & \multicolumn{3}{c|}{ hadr.}  \\
& & & [pb/$\gevsq$] & [pb/$\gevsq$]& [pb/$\gevsq$]& [pb/$\gevsq$]& [pb/$\gevsq$]&  \multicolumn{3}{c|}{ corr.}  \\
\hline
0.0 & - & 0.1 & 0.0   &-      & -     & -     & -     &      & -- &     \\
0.1 & - & 0.2 & 0.054 & 0.022 & 0.016 & 0.005 & 0.014 & 1.30 &$\pm$& 0.12\\
0.2 & - & 0.3 & 0.128 & 0.036 & 0.022 & 0.015 & 0.025 & 1.16 &$\pm$& 0.01 \\ 
0.3 & - & 0.4 & 0.160 & 0.036 & 0.023 & 0.011 & 0.025 & 1.16 &$\pm$& 0.05\\ 
0.4 & - & 0.5 & 0.150 & 0.039 & 0.023 & 0.018 & 0.026 & 1.16 &$\pm$& 0.03\\
0.5 & - & 0.6 & 0.105 & 0.024 & 0.018 & 0.010 & 0.011 & 1.14 &$\pm$& 0.01\\ 
0.6 & - & 0.7 & 0.075 & 0.016 & 0.013 & 0.004 & 0.006 & 1.13 &$\pm$& 0.04\\
0.7 & - & 0.8 & 0.058 & 0.014 & 0.012 & 0.005 & 0.005 & 1.05 &$\pm$& 0.28\\
0.8 & - & 0.9 & 0.052 & 0.015 & 0.013 & 0.005 & 0.005 & 1.01 &$\pm$& 0.21\\
0.9 & - & 1.0 & 0.025 & 0.015 & 0.013 & 0.005 & 0.005 &    &--&   \\
\hline
\end{tabular}
\caption{Bin averaged double differential cross sections of diffractive dijet production  at the hadron level
  (corrected to the  QED Born level) and the corresponding uncertainties as a function of $\zpom$ in different bins of $\mu^2=\qsq+p_{T, jet1}^2$.   The corrections applied to the NLO
  prediction for hadronisation and the associated uncertainty are also
  given. No hadronisation correction is given for the highest $\zpom$
  bin since it cannot be evaluated reliably.}
\label{data2}
\end{center}
\end{table}

%%% XS absolute in zpomeron %%%%%%%%%%%%%%%%%%%%%%%%%%%%%%%%%%%%%%%%

\end{document}